\documentclass[twocolumn,showpacs,preprintnumbers,amsmath,amssymb]{revtex4}
\usepackage{graphicx}
\usepackage{dcolumn}
\usepackage{bm}
\usepackage[tight]{subfigure}
\usepackage{soul,color}
\begin{document}
\title{Kinetic Equations for
  Transport Through Single-Molecule Transistors
}
\author{M. Leijnse$^{(1,2,3)}$}
\author{M. R. Wegewijs$^{(1,2,3)}$}
\affiliation{
  (1) Institut f\"ur Theoretische Physik A,
      RWTH Aachen, 52056 Aachen,  Germany \\
  (2) Institut f\"ur Festk{\"o}rper-Forschung - Theorie 3,
      Forschungszentrum J{\"u}lich, 52425 J{\"u}lich,  Germany \\
  (3) JARA- Fundamentals of Future Information Technology\\
}
\begin{abstract}
  We present explicit kinetic equations for quantum transport through a general molecular quantum-dot,
  accounting for all contributions up to 4th order perturbation theory in the tunneling Hamiltonian
  and the complete molecular density matrix.
  Such a full treatment describes not only sequential, cotunneling and pair tunneling,
  but also contains terms contributing to renormalization of the molecular resonances as well as their broadening.
  Due to the latter all terms in the perturbation expansion are automatically well-defined for any set of system parameters,
  no divergences occur and no by-hand regularization is required.
  Additionally we show that, in contrast to 2nd order perturbation theory,
  in 4th order it is essential to account for quantum coherence between \emph{non-degenerate} states,
  entering the theory through the non-diagonal elements of the density matrix.
  As a first application, we study 
  a single-molecule transistor coupled to a localized vibrational mode (Anderson-Holstein model).
  We find that cotunneling-assisted sequential tunneling processes involving the vibration
  give rise to \emph{current peaks} i.e. negative differential conductance in the Coulomb-blockade regime.
  Such peaks occur in the cross-over to strong electron-vibration coupling,
  where inelastic cotunneling competes with Franck-Condon suppressed 
  sequential tunneling, and thereby indicate the strength of the electron-vibration coupling.
  The peaks depend sensitively on the coupling to a
  dissipative bath, thus providing also an experimental probe of
  the Q-factor of the vibrational motion.
\end{abstract}
\maketitle
\section{Introduction.}
Electron transport through single-molecule transistors (SMTs) has been
intensively studied theoretically in recent years
\citep{Wegewijs05, Kaat05, Koch05b, Donarini06, Schultz07a, Reckermann08a, Reckermann08b, Koch04b, Koch06, Cornaglia05b, Koch05c, Flensberg03}
driven by ongoing experimental
advances~\citep{Yu04inel,Natelson06,Osorio07b,Parks07,Pasupathy04,Park00, Osorio07a}. One of the most distinctive features of SMTs, compared to artificially
nano-structured devices such as quantum dots, is the coupling between their \emph{quantized}
mechanical and electronic degrees of freedom~\cite{Osorio07a}.
The size and shape distortions of an SMT~\cite{Pasupathy04}  and its center-of-mass motion~\cite{Park00}
result in sharp transport resonances whose amplitudes are governed by the
quantum mechanical overlap of the corresponding \emph{mechanical} wavefunctions.
This {Franck-Condon} (FC) \emph{transport} effect is of fundamental interest
since it is induced by the change of molecular charge,
therefore involving strong electron charging and non-equilibrium effects,
in contrast to the usual FC-effect in optical spectroscopy where the charge remains unaltered.
The discrete vibrational modes of a molecule are also important in
assessing the atomistic details of the transport junction~\citep{Houck05}.
Finally, the demonstrated control over the molecular energy levels of an SMT using a gate electrode
provides interesting perspectives for realizing quantized nano-electromechanical systems (NEMS)\citep{Sapmaz05,Sapmaz03,Sazonova04}.
\par
The basic FC transport picture~\cite{Mitra04b} assumes single electrons to sequentially tunnel on and off the SMT.
This is valid in the limit of weak tunnel coupling and for applied gate- and bias-voltages such that
fluctuations of the molecular charge are not suppressed
by Coulomb interaction (Coulomb blockade) or quantum confinement effects.
In this limit it is sufficient to describe transport in lowest non-vanishing order perturbation theory in the tunneling
and many interesting results have been reported.
For instance, a well studied model in this context is the Anderson-Holstein model, consisting
of a spin-degenerate level with a linear coupling
(electron-vibration coupling, $\lambda$)
between the charge on the level and the coordinate of a vibrational mode.
When the overlap integrals between low lying vibrational states in two adjacent charge states of the SMT vanish,
a suppression of single-electron tunneling (SET) occurs, called Franck-Condon blockade~\citep{Koch04b}.
Here electron transport was found to take place through self-similar avalanches, leading to bunching of electrons and enhanced shot noise.
Extending the basic model with a charge-dependent vibrational frequency,
additional resonances occur~\citep{ Koch05b},
and interference of vibrational wavefunctions was shown to lead to a suppression of the electric current at finite bias~\citep{Wegewijs05}
due to a population inversion of the vibrational distribution.
More complex models with (quasi-)degenerate electronic orbitals and multiple modes exhibit (pseudo-)Jahn-Teller physics.
These may show rectifying behavior \citep{Kaat05},
dynamical symmetry breaking \citep{Donarini06} and
current suppression due to Berry phase effects~\citep{Schultz07a}.
Finally, distinctive transport signatures of the breakdown of the Born-Oppenheimer separation~\citep{Reckermann08a}
and correlations of vibration- and spin-properties have been predicted,
such as a vibration-induced spin-blockade~\cite{Reckermann08b}.
\par
Since the complicated transport processes in SMTs
may result in a suppression of single-electron tunneling,
it becomes more urgent to understand the effect of higher order tunnel processes.
Even more so since experimentally SMTs typically exhibit a significant tunnel coupling.
The purpose of this paper is to set up a general method to properly describe
\emph{all} coherent tunnel processes in leading and next-to-leading order in the tunneling Hamiltonian.
The method applies to very general molecular quantum dot models,
with many quantized excitations
and few relevant selection rules for transport quantities.
Additionally, all molecular interactions included in the model, for instance Coulomb charging and electron-vibration, are accounted for exactly
by from the outset formulating the transport equations in the basis of many-body eigenstates of the 
molecular Hamiltonian.
The non-linear transport is obtained using the explicitly calculated non-equilibrium density \emph{matrix}. 
A variety of next-to-leading order effects have been discussed previously.
For instance, a well known signature of higher order tunneling processes is the appearance of
inelastic cotunneling steps in the differential conductance, the position of which are
independent of the gate voltage, which have been observed in many experiments on semi-conductor
and molecular quantum-dots~\cite{Yu04inel,Natelson06,Parks07,Osorio07b}.
Additional gate-voltage dependent transport resonances have been found inside the Coulomb blockade
regime~\citep{Schleser05} and discussed theoretically~\citep{Golovach04, Elste07}. These resonances are due
to sequential tunneling events starting from states excited by inelastic cotunneling processes (``heating the molecule''),
called cotunneling-assisted sequential tunneling (COSET).
It was recently suggested~\citep{Luffe07} that these resonances can be used to experimentally estimate the decay-rate of vibrational excitations
due to a coupling to a dissipative environment.
Indeed, by extending the Golden-Rule approach by a next-to-leading order expansion of the T-matrix in the tunneling,
it was found~\citep{Koch06} that the COSET features are particularly pronounced in the Anderson-Holstein model in the limit
of large electron-vibration coupling.
Finally, effects of electron pair-tunneling were discussed~\citep{Koch05c} for an effective Anderson model with
attractive electron-electron interaction in the Golden-Rule approach using a Schrieffer-Wolff transformation.
\par
The method set up here captures all the aforementioned effects simultaneously.
By providing a microscopic derivation we overcome some drawbacks of the methods employed in the cited works,
related to accounting for broadening and renormalization of the molecular resonances, which were previously 
discussed in e.g.~\cite{Koenig98, Koenig99, Kubala06, Koch06},
see also Ref.~\cite{Timm08}.
The main focus of the paper is therefore on the general aspects of the transport theory.
As a central result we present explicit kinetic equations from which the full molecular density matrix
and transport current can be calculated.
We reformulate the real-time transport theory~\citep{Schoeller94, Koenig97} using Liouville super-operators
to present a straight-forward derivation.
The expressions for the transport rates are valid for a wide class of quantum-dot systems
and, importantly, involve no assumptions on model-specific selection rules.
We show that in such higher order calculations it is crucial
to include contributions from coherent superpositions of molecular states not protected by selection rules,
even when the level spacing is much \emph{larger} than the tunneling broadening.
The Anderson-Holstein model with a large vibrational
frequency compared to the tunneling coupling, $\hbar \omega \gg \Gamma$,
presents a case where this is extremely important 
and we demonstrate our method for this model.
This is in clear contrast to lowest order perturbation theory where these so-called
non-secular terms can be neglected.
We are not aware of previous works pointing this out.
\par
To maintain readability, the paper is divided into three parts: a general part~\ref{sec:Model}, application~\ref{sec:Anderson-Holstein} 
and technical Appendices.
In Sect.~\ref{sec:Model} we shortly describe the general model of a molecular quantum dot system
and the basic equations of the real-time transport theory.
We then discuss the central results,
the explicit transport equations 
for the full density matrix and transport current.
Detailed derivations and expressions are presented
in a coherent way in the Appendices for the theoretically interested reader.
In Sect.~\ref{sec:Anderson-Holstein} we study in detail the specific model of a molecular
transistor coupled to a localized vibrational mode, the non-equilibrium Anderson-Holstein model.
We summarize and conclude in Sect.~\ref{sec:Conclusions}.
\par
Throughout the rest of the paper we use natural units where $\hbar = k_\text{B} =|e| = 1$ where $-|e|$ is the electron charge.

\section{
  Model and transport theory
  \label{sec:Model}}
We consider a molecule as a complex quantum dot, connected to a number of macroscopic reservoirs labeled by $r$.
The electrons in the reservoirs are considered to be non-interacting, but no assumptions
are made concerning the type or strength of the interactions on the molecule, as long as we can
diagonalize the isolated molecular many-body Hamiltonian.
The entire system is described by the Hamiltonian
$H_{\text{tot}}=H + H_\text{R} + H_\text{T}$, where $ H_\text{R} = \sum_r H_r$ and
\begin{eqnarray}
  \label{eq:ham_mol}
  H    &=& \sum_{a} E_a |a \rangle \langle a |, \\
  \label{eq:ham_res}
   H_r &=& \sum_{\sigma} \int d \omega ~\omega c_{r \sigma -\omega} c_{r \sigma +\omega}, \\
  \label{eq:ham_tun}
   H_\text{T} &=&  \sum_{r \sigma N} \sum_{\eta = \pm 1} \sum_{a \in N}^{a' \in ({N - \eta})} \eta \int d \omega
                   ~T_{r \sigma \eta}^{a a'} |a \rangle \langle a' | c_{r \sigma \eta \omega}.
\end{eqnarray}
The Hamiltonians are written from the outset in a form which deviates from that commonly used.
This allows crucial simplifications of the derivations and explicit expressions presented in the Appendices.
In the molecular Hamiltonian, $H$, $|a \rangle$ denotes a general many-body eigenstate with energy $E_a$.
We assume that we can classify these states by the number of excess electrons, $N$, on the molecule.
The electron number, together with other quantum numbers depending on the model at hand
(e.g. spin, magnetic and vibrational quantum numbers), are labeled by $a$.
We will loosely denote
by $N_a$ the electron number in state $a$.
$H_r$ describes reservoir $r$ and is written in terms of continuum field operators
\begin{eqnarray}
  \label{eq:annihilation}
  c_{r \sigma + \omega} &=& \sum_k \frac{1}{\sqrt{\rho_{r \sigma}}} \delta( \epsilon_{r \sigma k} - \omega) c_{r \sigma k}, ~~\eta = +, \\
  \label{eq:creation}
  c_{r \sigma - \omega} &=& \sum_k \frac{1}{\sqrt{\rho_{r \sigma}}} \delta( \epsilon_{r \sigma k} - \omega) c_{r \sigma k}^{\dagger}, ~~\eta = -,
\end{eqnarray}
where $c_{r \sigma k}^{\dagger}$ ($c_{r \sigma k}$) are the usual creation (annihilation) operators for electrons
in reservoir $r$ with spin-projection $\sigma$, state-index $k$ and energy $\epsilon_{r \sigma k}$.
We will refer to $\eta$ as the electron-hole (e-h) index.
$\rho_{r \sigma}$ is the density of states.
Inserting~(\ref{eq:annihilation}-\ref{eq:creation}) into~(\ref{eq:ham_res}) one recovers the standard form
of the reservoir Hamiltonian. For this one assumes that there is a unique correspondence between $k$ and $\epsilon_{r \sigma k}$.
For cases where this does not hold, one labels different branches of the dispersion relation by an additional index.
The tunnel Hamiltonian $H_{\text{T}}$ describes the tunneling into or out of the molecule,
involving a change of the molecular state from $a'$ to $a$.
The relevant matrix elements are given by superpositions of single-particle tunneling matrix elements $t_{l r \sigma}$
and many-body amplitudes of the molecular wavefunction:
\begin{eqnarray}
  \label{eq:TMEplus}
  T_{r \sigma +}^{a a'} &=& \sqrt{\rho_{r \sigma}} \sum_l t_{l r \sigma} \langle a | d_{l \sigma}^{\dagger} | a' \rangle, \\
  \label{eq:TMEminus}
  T_{r \sigma -}^{a a'} &=& \sqrt{\rho_{r \sigma}} \sum_l t^{*}_{l r \sigma} \langle a | d_{l \sigma} | a' \rangle
  			 = \left( T_{r \sigma +}^{a' a} \right)^{*}.
\end{eqnarray}
Here $l$ labels a single particle basis for the molecule 
with corresponding creation / annihilation operator $d_{l \sigma}^{\dagger}, d_{l \sigma}$.
Note that the density of states is incorporated in $T_{r \sigma \eta}^{a a'}$, simplifying many expressions.
The spectral densities
$\Gamma^{a b, a' b'}_{r \sigma} = 2 \pi  T_{r \sigma +}^{a a'} T_{r \sigma -}^{b' b}$,
thus represent the set of relevant energy scales for the tunneling.
Both $\rho_{r\sigma}$ and $t_{l r \sigma}$ are assumed to be energy independent.
This is the most relevant physical limit and presents no principle limitation of the presented method (only numerical).
Charge conservation implies the selection rule, $N_a - N_{a'} = 1$,
which is contained in
$\langle a | d_{l \sigma}^{\dagger} | a' \rangle \propto \delta_{N_a, N_{a'} + 1}$.
This is the only selection rule assumed.
A Fermion sign, $\eta$, appears in Eq.~(\ref{eq:ham_tun}) since we always write the reservoir operator
to the right of the projector. However, one can show that this exactly cancels in all expressions involving an average
over the reservoir degrees of freedom. It cancels with an extra Fermion sign appearing 
when disentangling the dot and reservoir operators, using that an equal number of creation and annihilation operators
must occur to give a non-zero average, see Ref.~\cite{Schoeller08-rtrg} for a proof.
We can therefore discard the sign $\eta$ from the outset, and everywhere treat dot and lead 
operators as commuting, greatly simplifying the calculation of signs.
\subsection{Kinetic Equation}\label{sec:RT}
\par
A microscopic molecular system coupled to macroscopic reservoirs is
completely described by its reduced density operator $P(t)$,
obtained by averaging the total density operator over the reservoir degrees of freedom,
$P(t) = \text{Tr}_\text{R} \rho(t)$.
The reduced density operator evolves in time according to a quantum kinetic equation.
The presence of strong non-equilibrium effects (non-linear transport) and strong local interactions
(Coulomb, electron-vibration, etc.) makes the calculation of the transport rates occurring in this equation a cumbersome task.
Here our goal is to derive explicit expressions for the next-to-leading order transport rates
in terms of the parameters $E_a, T^{a a'}_{r \sigma \eta}$ of 
the Hamiltonians~(\ref{eq:ham_mol}--\ref{eq:ham_tun}) and the statistical properties 
of the electrodes $T$ (temperature) and $\mu_r$ (chemical potential).
The real-time transport theory, developed in~\citep{Schoeller94, Koenig97}
and extended by several groups~\citep{Thielmann05, Weymann05, Kubala06},
provides straightforward rules for the calculation of the transport rates
using a diagrammatic representation on the Keldysh contour,
avoiding any spurious regularization problems.
This technique has been simplified further by using special Liouville super-operators
and corresponding diagrams in the context of a non-equilibrium renormalization group approach~\cite{Schoeller08-rtrg, Korb07}.
For clarity we discuss the general aspects here, whereas the important but
cumbersome expressions are coherently derived and presented in the Appendices.
The starting point is the time evolution of the density operator of the total system, molecule + reservoirs:
\begin{eqnarray}
  \label{eq:total_rho}
	\rho(t) = e^{-i L_{\text{tot}} t} \rho(0).
\end{eqnarray}
Here the Liouvillian super-operators in $L_{\text{tot}} = L + L_\text{R} + L_\text{T}$,
act on an arbitrary operator $A$ by forming a commutator with the Hamiltonian,
e.g. $L A = [H, A]$.
We assume the system to be decoupled at the initial time $t =0$, such that the density operator
factorizes, $\rho(0) = P(0) \rho_\text{R}$ where
$\rho_\text{R} = \prod_r \rho_r$ and $\rho_r$ describes reservoir $r$.
Each reservoir is assumed to remain in internal equilibrium independently
and is described by a grand-canonical ensemble at all times.
When a bias voltage is applied to the system, causing the chemical potentials of different leads to differ, 
this puts an inhomogeneous ``boundary condition'' on the molecular density operator 
and drives it out of equilibrium.
We now take the Laplace transform of Eq.~(\ref{eq:total_rho}) and trace
out the reservoirs
\begin{eqnarray}
  \label{eq:laplace_rho1}
          P(z) &=& \underset{\text{R}}{\text{Tr}} \int_0^{\infty} d t e^{i z t} e^{-i L_{\text{tot}} t} P(0) \rho_\text{R} \\
  \label{eq:laplace_rho2}
	       &=& i \underset{\text{R}}{\text{Tr}} \frac{1}{z - L_\text{R} - L - L_\text{T}} P(0) \rho_\text{R} \\
  \label{eq:laplace_rho3}
  	       &=& \frac{i}{z - L - iW(z)} P(0),
\end{eqnarray}
where the last expression is obtained by expanding the denominator in~(\ref{eq:laplace_rho2})
in powers of the tunneling Liouvillian, $L_\text{T}$, carrying out the trace over the reservoirs
and re-summing the series, see Appendix~\ref{sec:kinetic_equation} for details.
Here $iW(z)$ is a (super-operator) self-energy and $L + iW(z)$
describes the molecular density operator in the presence of the reservoirs.
If Eq.~(\ref{eq:laplace_rho3}) is transformed back to the time-domain, $W(t - t')$ appears as a
kernel in the integro-differential equation for $P(t)$
\begin{eqnarray}
  \label{eq:kinetic_equation_timespace}
  \dot P(t) = -i L P(t) + \int_0^t d t' W(t - t') P(t'),
\end{eqnarray} 
assuming that the Laplace transform of $W(t-t')$ exists.
We are exclusively interested in the stationary state at $t\rightarrow \infty$ 
(asymptotic solution) of the molecular density operator,
i.e. the zero frequency limit $z \rightarrow i0$
where the imaginary infinitesimal physically originates from the adiabatic switching on of the tunneling.
Assuming that a unique stationary state exists and using
$ \text{lim}_{t \rightarrow \infty} P(t)=-i \text{lim}_{z \rightarrow i0} z P(z)$,
Eq.~(\ref{eq:laplace_rho3}) gives the 
standard form~\cite{Schoeller94, Koenig99, Schoeller08-rtrg} of the
stationary state equation:
\begin{eqnarray}
  \label{eq:kinetic_equation}
          0 = \left( -i L + W \right) P.
\end{eqnarray}
Here, and in the rest of the paper, we use the notation $W = \text{lim}_{z \rightarrow i0} W(z)$
and $P = \text{lim}_{t \rightarrow \infty} P(t)$.
Supplemented with the probability normalization condition $\text{Tr}_\text{M} P = 1$,
where $\text{Tr}_\text{M}$ is the trace over the molecular degrees of freedom,
this uniquely determines the stationary state.
The normalizability derives from the general property of the kernel 
$\text{Tr}_{\text{M}} W A = 0$ for any operator $A$.
Matrix elements of a super-operator, $S$, are defined according to
\begin{eqnarray}
  \label{eq:supermatrix-element}
	S^{a' b'}_{a b} \equiv \langle a | \left( S |a'\rangle \langle b' | \right) | b \rangle,
\end{eqnarray}
meaning that we first act with $S$ on a projector $|a'\rangle \langle b' |$, generating
a new operator, and subsequently take matrix elements of this.
In the basis of the many-body eigenstates of the isolated molecule the molecular Liouvillian
is
\begin{eqnarray}
\label{eq:LM_matrix_elements}
		L_{a b}^{a' b'} = (E_a - E_b) \delta_{a a'} \delta_{b b'}.
\end{eqnarray}
Our main objective is to calculate the expectation value of the electron current flowing out of 
reservoir $r$ into the molecule, $I_r(t) = \text{Tr}\hat{I}_r \rho(t)$ where $\text{Tr}$ is the trace over the full system
and $\hat{I}_r = -\frac{d}{d t} N_r = -i[H_\text{T}, N_r]$, with $N_r$ being the number operator for 
electrons in reservoir $r$. As is shown in
Appendix~\ref{sec:kinetic_equation}, this expectation value
can be obtained from a kernel similar to $W$ and the density operator.
In the stationary state: $I_r = \text{Tr}_{\text{M}} \{ W_{I_r} P \}$, where the current kernel,
$W_{I_r}$, contains the subset of tunneling processes described by $W$
which contribute to the current through reservoir $r$.
We can now write down the generalized, formally exact, master equations
\begin{eqnarray}
  \label{eq:master_equation_rho}
	0  &=& \sum_{a' b'} \left[ -i L_{a b}^{a' b'}
		+ \sum_{k = 1}^{\infty} \left( W^{(2k)} \right)_{a b}^{a' b'} \right] P_{a' b'}, \\
  \label{eq:master_equation_norm}
	1 &=& \sum_{a} P_{a a}, \\
  \label{eq:master_equation_I}
	I_r &=& \sum_{a} \sum_{a' b'} \sum_{k} \left( W_{I_r}^{(2k)} \right)^{a' b'}_{a a} P_{a' b'}.
\end{eqnarray}
Here we have expanded the kernels in even order terms $2k$ in the tunneling Liouvillian
accounting for coherent $k$-electron tunnel processes
(odd orders vanish when tracing over the reservoirs since the tunneling Hamiltonian is linear in reservoir field operators).
Eq.~(\ref{eq:master_equation_rho}--\ref{eq:master_equation_I}) compactly formulate the transport problem. The first 
central result of this paper is the explicit evaluation of the kernels $W^{(2)}$ and $W^{(4)}$
as given in Appendix~\ref{sec:2ndorder} (Eq.~(\ref{eq:W2})) and \ref{sec:4thorder} (Eq.~(\ref{eq:W4explicit})),
accounting for coherent single- and two-electron tunneling processes.
Their detailed form is not needed here, but an important property of these expressions
is that they are finite by construction for any system
parameters and applied voltages at non-zero temperature:
no divergences occur and no by-hand regularization is required at any stage of the calculation
as is the case in the Golden-Rule T-matrix approach~\cite{Koch06}.
Of course, the finite temperature must be chosen sufficiently large compared to
the tunneling couplings ($\Gamma \ll T$) to avoid the breakdown of perturbation theory.
\par
For the solution of the kinetic equation it is important to know whether
the molecular density matrix is diagonal in certain quantum numbers due to a conservation law.
The only such law explicitly enforced here concerns the total charge in reservoirs + molecule:
$[H_\text{tot}, N_\text{tot}]=0$.
As is shown in Appendix~\ref{sec:diagram_rules},
the matrix elements of $W^{a' b'}_{a b}$ vanish unless the charge differences are equal:
$N_{a'} - N_{b'} = N_a - N_b$.
With the assumption that the density matrix is diagonal with respect to charge at $t=0$, before
the coupling to the reservoirs is switched on, it is guaranteed to remain so at all times.
In a similar way any conserved quantity of the \emph{total} system
encodes selection rules in the tunneling matrix elements ensuring that
the density matrix remains diagonal in the corresponding molecular quantum number.
For example, for the Anderson-Holstein model studied in Sect.~\ref{sec:Anderson-Holstein} the conservation of total spin-projection,
 $S_z$, leads to a density matrix which is diagonal in the spin-projection of the molecule, $s_z$.
\subsection{Solution of the Kinetic Equation}
The solution of equations~(\ref{eq:master_equation_rho}-\ref{eq:master_equation_I})
with perturbatively calculated kernels (up to a finite order) for the \emph{full} molecular density matrix requires some care
for models with excited states and tunnel matrix elements without strict selection rules.
The present section is therefore devoted to deriving the correct and well-behaved master equations in 
next-to-leading order perturbation theory.
First we rewrite the equations by collecting the elements of the density operator into a vector, $\mathbf{P}$,
and the elements of the rate super-operators into matrices $\mathbf{W}, \mathbf{W}_{I_r}, \mathbf{L}$ acting on this vector.
Up to 4th order in the perturbation expansion the equations can now be written as
\begin{eqnarray}
        \label{eq:crossover_scheme_p}
        \mathbf{0} &=& \left(-i \mathbf{L} + \mathbf{W}^{(2)} + \mathbf{W}^{(4)} \right) \mathbf{P},
        \\
        \label{eq:crossover_scheme_norm}
        1 &=& \mathbf{e}^T \mathbf{P},
        \\
        \label{eq:crossover_scheme_I}
	I_r &=& \mathbf{e}^T \left( \mathbf{W}_{I_r}^{(2)} + \mathbf{W}_{I_r}^{(4)} \right) \mathbf{P}.
\end{eqnarray}
The trace in Eq.~(\ref{eq:master_equation_norm}), (\ref{eq:master_equation_I})
is effected by the multiplication with the auxiliary vector $\mathbf{e}^T = (1,\ldots,1,0,\ldots,0)$
to sum up all vector elements corresponding to diagonal density-matrix elements.
The sum-rule on the kernel reads $\mathbf{e}^T \mathbf{W}=\mathbf{0}^T$.
\subsubsection{Elimination of Non-Diagonal Elements}
The crucial assumption for the following discussion is that the spectrum is free
from accidental degeneracies in the following sense:
all pairs of states $a \neq b$ which have non-zero non-diagonal density matrix elements $P_{ab}$
are well separated in energy on the scale set by the tunneling rates.
Models for molecular transistors with discrete vibrational modes, such as the Anderson-Holstein model,
satisfy this condition, provided that the vibrational level-spacing is larger than the tunneling coupling,
since spin-selection rules
generally prohibit coherence between the degenerate spin-states, unless broken 
by e.g. magnetic anisotropy or spin-polarization of the electrodes. 
One can always eliminate the non-diagonal elements and incorporate their effect in a
correction to the rates coupling diagonal elements.
To this end we collect diagonal ($d$) and non-diagonal ($n$) density matrix elements 
into separate vectors $\mathbf{P}_d$ and $\mathbf{P}_n$, separate~(\ref{eq:crossover_scheme_p}) into
blocks and denote $\mathbf{W} = \mathbf{W}^{(2)} + \mathbf{W}^{(4)}$:
\begin{eqnarray}
  \label{eq:block_matrices}
	\left[
	\begin{array}{c}
		\mathbf{0}_d \\
		\mathbf{0}_n
	\end{array}
	\right] =
	\left[
	\begin{array}{cc}
		\mathbf{W}_{d d} & \mathbf{W}_{d n} \\
		\mathbf{W}_{n d} & \mathbf{W}_{n n} - i \mathbf{L}_{nn}
	\end{array}
	\right]
	\left[
	\begin{array}{c}
		\mathbf{P}_d \\
		\mathbf{P}_n
	\end{array}
	\right].
\end{eqnarray}
It is clear from~(\ref{eq:LM_matrix_elements}) that $\mathbf{L}$ is only non-zero in the $n n$ block.
The sum-rule implies
\begin{eqnarray}
       \label{eq:sumrule_dd}
        \mathbf{e}_d^T \mathbf{W}_{d d} & = & \mathbf{0}_d^T,
        \\
       \label{eq:sumrule_dn}
        \mathbf{e}_d^T \mathbf{W}_{d n} & = & \mathbf{0}_n^T,
\end{eqnarray}
where the multiplication with the vector $\mathbf{e}^T_d = (1,\ldots,1)$ sums up all $d$-vector elements.
We can now eliminate processes into the non-diagonal sector of the density-matrix
by solving the equation from the lower block for the non-diagonal part of the density matrix, $\mathbf{P}_n$,
and inserting this back into the equation in the upper block for the diagonal part.
Due to the clear separation of energy scales (non-degenerate spectrum) we can expand in the small quantity 
$\mathbf{W}_{n n} \mathbf{L}_{nn}^{-1}$.
Consistently neglecting terms of order
$> H_{\text{T}}^{4}$ we then obtain an effective equation for $\mathbf{P}_d$:
\begin{eqnarray}
  \label{eq:effective_rate_approx}
                \mathbf{W}_{d} \mathbf{P}_d  &=& 0,
                \\
                \label{eq:effective_norm}
                \mathbf{e}_{d}^T \mathbf{P}_d  &=& 1,
\end{eqnarray}
\begin{eqnarray}
\label{eq:effective_rate}
		\mathbf{W}_{d} &=&
                \mathbf{W}_{dd}^{(2)} + \mathbf{W}_{dd}^{(4)} -
		i \mathbf{W}_{dn}^{(2)} \mathbf{L}_{nn}^{-1} \mathbf{W}_{nd}^{(2)}.
\end{eqnarray}
The diagonal elements of the density matrix (vector of probabilities) thus satisfy what looks like a classical
rate-equation, but with the \emph{effective rates}~(\ref{eq:effective_rate}).
A completely analogous calculation for the correction to the current
from non-diagonal elements gives
\begin{eqnarray}
  \label{eq:effective_current_expansion}
                I_r &=& \mathbf{e}^T_d \left( \mathbf{W}_{I_r} \right)_{d} \mathbf{P}_d,
\end{eqnarray}
\begin{eqnarray}
  \label{eq:effective_current}
                 & \left( \mathbf{W}_{I_r} \right)_{d} = & \nonumber \\
                & \left( \mathbf{W}_{I_r}^{(2)}\right)_{dd} +
                \left( \mathbf{W}_{I_r}^{(4)}\right)_{dd}
                - i \left( \mathbf{W}_{I_r}^{(2)}\right)_{dn} \mathbf{L}_{nn}^{-1} \mathbf{W}_{nd}^{(2)}.
                &
\end{eqnarray}
It can easily be shown that~(\ref{eq:effective_rate}) and~(\ref{eq:effective_current})
are real, ensuring that the diagonal elements of the density matrix as well as the current are real.
Due to Eq.~(\ref{eq:sumrule_dd}--\ref{eq:sumrule_dn}) the effective rate matrix satisfies the sumrule
$\mathbf{e}_{d}^T\mathbf{W}_{d} =\mathbf{0}^T$,
so that Eq.~(\ref{eq:effective_rate_approx}) with Eq.~(\ref{eq:effective_norm}) determine 
the unique stationary solution for the vector of diagonal density matrix elements (probabilities).
Eq.~(\ref{eq:effective_rate_approx}-\ref{eq:effective_current}) form
another central result of this work and we comment on their significance and importance.
The advantage of the formulation in terms of effective rates, compared to solving 
Eq.~(\ref{eq:crossover_scheme_p}--\ref{eq:crossover_scheme_I}) directly, is 
threefold:
(i) the effective rate matrices include the effects of coherences only up to order $H_\text{T}^4$,
just as the other effects of tunneling;
(ii) it makes it explicit that the 2nd order coherences effectively give 4th order effects in the
rates for the occupations, something which is hidden in Eq.~(\ref{eq:crossover_scheme_p});
(iii) it shows that the large matrix $\mathbf{W}_{nn}$, 
as well as all 4th order matrices which are not diagonal in initial and final state indices, need not be evaluated,
significantly simplifying the calculation.
The appearance of the correction in the effective rate has an intuitive physical meaning in the time-domain:
it corresponds to a process starting ($\mathbf{W}_{nd}$) and ending ($\mathbf{W}_{dn}$) in a diagonal state, through two tunnel processes.
In the intermediate non-diagonal state the free evolution involves rapid coherent oscillations at the Bohr-frequencies 
contained in $\mathbf{L}_{nn}$ (see~(\ref{eq:LM_matrix_elements})).
Due to the latter, these so-called non-secular terms~\cite{Blum_book} should be neglected
in a lowest order approximation.
However, these correction terms from coherences between \emph{non-degenerate states},
although formally containing only 2nd order rates, contribute in 4th order to the occupancies,
where they are crucial unless special model properties (conservation laws) make the matrix $\mathbf{W}_{dn}$ vanish exactly.
They scale in the same way as processes described by $\mathbf{W}_{dd}^{(4)}$
when one uniformly reduces the tunneling matrix elements.
Finally, we have found by numerical calculations for several model systems
that partial cancellations between the non-diagonal correction terms
and diagonal 4th order terms are crucial for obtaining a physical result:
if these corrections are excluded one obtains SET-like resonances in the Coulomb-blockade regime
\emph{below} the inelastic cotunneling threshold.
These are artifacts due to incorrect, large occupations of the excited states, even at zero bias voltage.
Depending on the parameters of the model, negative occupation probabilities may even result,
particularly when the tunneling amplitudes~(\ref{eq:TMEplus}-\ref{eq:TMEminus}) vary strongly from state to state.
Accounting for the non-diagonal correction terms no such artifacts occur.
Models for SMTs are typical systems where the neglect of these non-diagonal corrections
results in dramatic, spurious effects in the occupations and current.
\par
Summarizing: in the limit of large level-spacing,
Eq.~(\ref{eq:effective_rate_approx}--\ref{eq:effective_current})
are the correct expressions for the occupation probabilities and current.
The corrections from 2nd order non-diagonal terms
contribute only in 4th order in $H_{T}$:
in a consistent 2nd order calculation they must be omitted
whereas in a 4th order calculation they must be kept
unless all non-diagonal elements vanish due to selection rules.
\subsubsection{Calculation of Diagonal Elements}
\begin{figure}[t!]
  \hfill
  \begin{minipage}[l!]{.45\textwidth}
    \begin{center}
     	\includegraphics[height=0.69\linewidth]{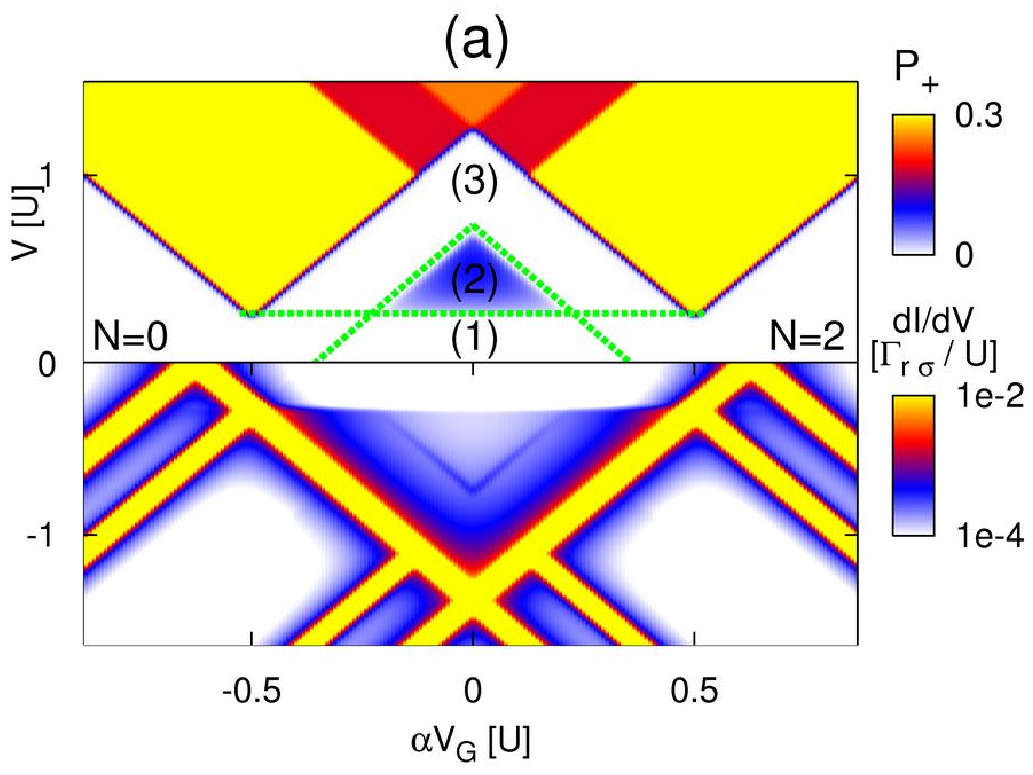}
    \end{center}
  \end{minipage}
  \hfill
  \begin{minipage}[l!]{.45\textwidth}
    \begin{center}
      \includegraphics[height=0.2\linewidth]{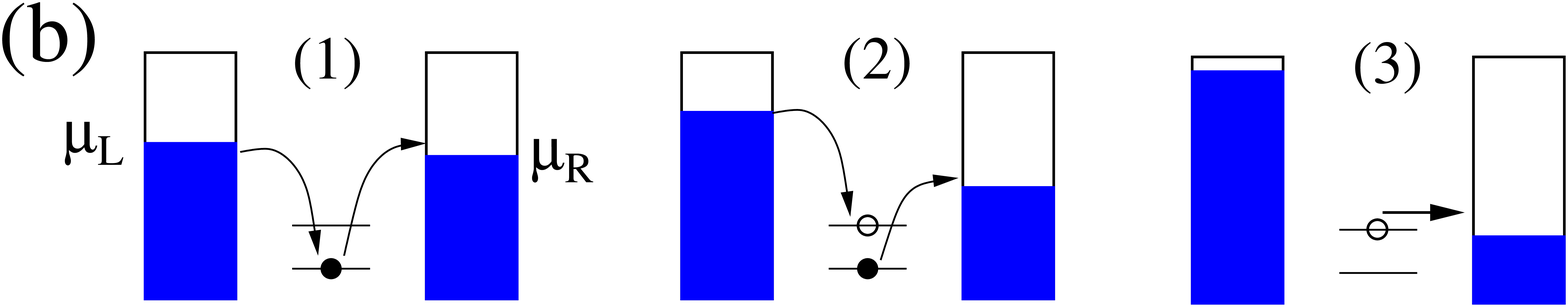}
    \end{center}
  \end{minipage}
  \hfill
  \caption{
    \label{fig:1}
     (Color online). (a): Occupation of excited spin-state, $P_+$ (top, positive bias) and differential
     conductance (bottom, negative bias) for the single-level Anderson model where the spin degeneracy is
     lifted by an applied magnetic field.
     Here $\Gamma_{\text{L} \sigma} = \Gamma_{\text{R} \sigma} = 10^{-2}T = 5 \times 10^{-5} U$, where $U$ is the charging energy, and
     $\epsilon_{\uparrow} - \epsilon_{\downarrow} = 50T$.
     (b): Energy diagrams in the regions (1), (2) and (3), separated by green dashed lines in (a).
     In (1) only elastic cotunneling is energetically possible.
     In (2) also inelastic cotunneling can take
     place, but the only way for the excited system to return to the ground state is by another inelastic
     cotunneling process.
     In (3) the excited state can be emptied also by sequential tunneling processes (COSET).
}
\end{figure}

Having eliminated the non-diagonal elements,
the remaining problem is the solution of the kinetic equation for the diagonal 
elements~(\ref{eq:effective_rate_approx}--\ref{eq:effective_norm}) only.
This requires some care since the effective rates~(\ref{eq:effective_rate}) 
contain both 2nd and 4th order terms,
as was discussed in previous works~\citep{Weymann05, Becker07} (where corrections from 
non-diagonal elements were exactly zero due to selection rules).
The problem is most easily understood from a simple example.
Fig.~\ref{fig:1}(a)
shows the result of 4th order perturbation theory for the single-level 
Anderson model in a magnetic field by solving
Eq.~(\ref{eq:effective_rate_approx}--\ref{eq:effective_current})
(due to spin-conservation $\mathbf{W}_{dn}^{(2)}=\mathbf{0}$ and non-diagonal elements play no role)
and Fig.~\ref{fig:1}(b) indicates relevant tunneling processes in regions (1), (2) and (3) in (a).
In region (2) the excited spin-state can be populated by 4th order processes (inelastic cotunneling).
Since sequential tunneling out of this state is only possible at larger bias (region (3)), it
can only relax by another inelastic \emph{cotunneling} process back into the ground state.
This latter process would not be included if one insists on an order-by-order solution, 
i.e. expand also the occupation vector in powers of $H_\text{T}$:
$ \mathbf{P}= \delta \mathbf{P}^{(0)}+ \delta \mathbf{P}^{(2)}$, solve for 
$\delta \mathbf{P}^{(0)}$ and $\delta \mathbf{P}^{(2)}$ separately and discard the term
$ \mathbf{W}^{(4)}_{d d} \delta \mathbf{P}^{(2)}$ in~(\ref{eq:effective_rate_approx}), which is formally of order 6.
Such an approach thus breaks down since the excited state is ``pumped'' up by inelastic tunneling, 
but not allowed to relax by 4th order processes, yielding an unphysical solution:
{$ \mathbf{W}^{(4)}_{d d} \delta \mathbf{P}^{(0)}$ provides an inflow into the excited spin-state, 
but no outflow since $\delta \mathbf{P}^{(0)}$ is only finite for the ground state, 
resulting in an artificially large correction $\delta \mathbf{P}^{(2)}$}.
Eq.~(\ref{eq:effective_rate_approx}) on the other hand has well behaved solutions, in which the occupancy of the excited spin-state
is determined by the competition of in- and out-going 4th order rates.
We emphasize that the problem with the order-by-order solution is not of a numerical nature and occurs even if 
the equations are solved analytically.
It is of a general nature and occurs whenever all lowest order rates connected to some state 
are suppressed. 
Ref.~\citep{Weymann05} suggests dividing the Coulomb diamond into different regions, adapting the 
solution of the master equations thereafter, e.g. using the order-by-order solution in the 
SET regime only.
However, for a general SMT model such a division is not possible since even in the SET regime some 
rates may be suppressed by e.g. Franck-Condon or magnetic blockade effects. 
Always solving Eq.~(\ref{eq:effective_rate_approx}) guarantees a physical solution in the sense that 
in- and out-going rates of all states are treated on an equal footing and the accuracy of the method is 
only limited by the order of the perturbation expansion of the kernel $W$.
\section{Non-equilibrium Anderson-Holstein model}\label{sec:Anderson-Holstein}
We now turn our focus to the Anderson-Holstein model, choosing the specific form 
of the molecular Hamiltonian~(\ref{eq:ham_mol})
\begin{eqnarray}
  \label{eq:ham_AH}
	H = \epsilon \sum_{\sigma} d^{\dagger}_{\sigma} d_{\sigma} + \frac{U}{2} \hat n (\hat n - 1)
	+ \omega (b^{\dagger} b + \frac{1}{2}).
\end{eqnarray}
The first two terms describe an electron in a single molecular orbital
with electron operators $d^{\dagger}_{\sigma},d_{\sigma}$ for spin $\sigma$ and $\hat n = \sum_{\sigma}
d^{\dagger}_{\sigma} d_{\sigma}$ denotes the number of excess electrons on the SMT.
The last term describes the quantized vibration of the SMT through the operators $b^{\dagger},b$.
The eigenstates $|a \rangle$ in Eq.~(\ref{eq:ham_mol}) thus have an electronic and a vibrational
part, $|a \rangle = |e \rangle | m_e \rangle$,
where $|e \rangle = |0 \rangle$, $|\uparrow \rangle$, $|\downarrow \rangle$, $|\uparrow \downarrow \rangle$
denotes the electronic state with $N = 0, 1, 2$ excess electrons on the molecule, and $|m_e \rangle$ labels the
state of the oscillator.

\begin{figure}
   \includegraphics[scale=0.15]{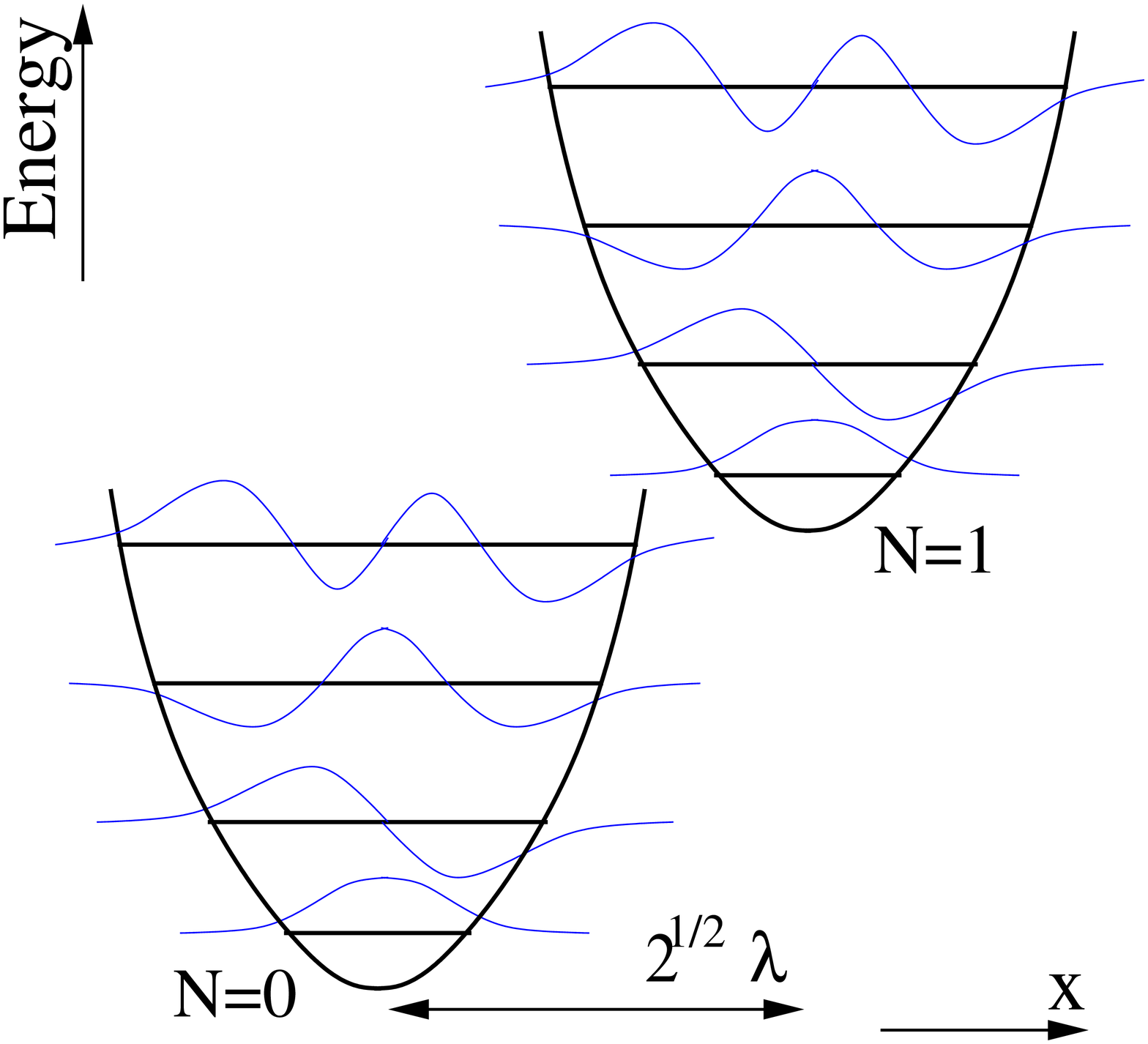}
   \caption{
     \label{fig:2}
   (Color online).
   Vibrational potentials in charge state $N = 0$ and $N = 1$ and lowest
   corresponding vibrational wavefunctions.
   The minimum of the potentials are shifted along the vibrational coordinate $x$ by $\sqrt{2} \lambda$ 
   (in units of the zero-point amplitude of the oscillator).
   }
\end{figure}

We have written Eq.~(\ref{eq:ham_AH}) in the standard polaron-basis where $\epsilon$
denotes the experimentally controllable effective energy level and $U$ denotes the effective charging energy,
both containing polaron-shift corrections~\cite{Cornaglia05b}.
The dimensionless electron-vibration coupling, denoted by $\lambda$,
appears in an operator which displaces the 
vibrational states by $\sqrt{2}\lambda$ 
along the vibrational coordinate (normalized to the zero-point amplitude), 
whenever an electron tunnels from the electrodes onto the molecule, see Fig.~\ref{fig:2}.
Thus the addition of an electron to the SMT induces a transition
$N\rightarrow N+1$, accompanied by a change of its vibrational state $m' \rightarrow m$.
The matrix element for this process is reduced
relative to the pure electronic tunneling amplitude by the
Franck-Condon overlap of vibrational wavefunctions in two different charge states of the SMT:
\begin{eqnarray}
\label{eq:fc-amplitude}
 f_{m m'} &=& \langle m | e^{-\lambda(b^{\dagger} - b)} | m' \rangle \\
 	  &=& \left( - \lambda \right)^{m - m'} 
	  e^{-\frac{\lambda^2}{2}} \sqrt{\frac{m'!}{m!}} L_{m'}^{m - m'}(\lambda^2),\nonumber 
\end{eqnarray}
for $m \geq m'$ (replace $m \leftrightarrow m'$ for $m < m'$)
where $L_{j}^{i}(x)$ is the generalized Laguerre polynomial.
The dependence of the Franck-Condon factors on $m$ and $m'$ was discussed in detail in~\citep{Wegewijs05}.
For a tunneling event starting from the vibrational ground state, $m'=0$, the Franck-Condon factors have a significant amplitude
for $\lambda^2 - \lambda \lesssim m \lesssim \lambda^2 + \lambda$, corresponding to classically allowed transitions.
More generally, for moderate to strong coupling there are broad regions in the $m,m'$ plane, bounded by the 
so-called Franck-Condon parabola, where vibration assisted transitions have significant amplitude.
These finite amplitudes for transitions to a range of vibrational states make the coherence between all pairs 
of these states important for the 4th order calculation, even though they are non-degenerate
(on the scale of the tunneling broadening), i.e. there are many non-zero elements of $\mathbf{W}_{dn}$ and $\mathbf{W}_{nd}$ 
in~(\ref{eq:effective_rate}).
\par
Here we are interested in transport close to a charge degeneracy point,
accounting for the fact that the charging energy together with the
confinement-induced level-spacing typically constitute the largest energy scales in SMTs.
We therefore restrict the model to electronic states with charge $N=0,1$, 
equivalent to taking $U \rightarrow \infty$ in Eq.~(\ref{eq:ham_AH}).
Without loss of generality we take $\epsilon = -\alpha V_\text{G}$, where $\alpha$ is the gate-coupling factor, i.e.
we associate $\epsilon=0$ with zero gate voltage.
The tunneling matrix elements for an electron tunneling onto the molecule are given by 
$T_{r \sigma +}^{a a'} = \delta_{s_z \sigma} \sqrt{\rho} t_{r} f_{m m'}$,
where the eigenstates are labeled by the quantum numbers $a = (s_z, m)$ in the $N=1$ charge state and
$a' =(0, m')$ in the $N=0$ charge state, with $s_z$ denoting the spin-projection of the molecule. 
We have everywhere used $\omega = 40 T = 10^4 \Gamma_\text{M}$, where $\Gamma_\text{M}$ is the maximum sequential
tunneling rate, i.e. $\Gamma_\text{M} = \Gamma \times \text{max}(|f_{m m'}|^2)$
and $\Gamma = |2 \pi \sqrt{\rho} t_L|^2 = |2 \pi \sqrt{\rho} t_R|^2$
is the pure electronic tunneling rate for symmetric coupling to the left and right electrodes.
We set the width of the conduction band to $D = 250 \omega$.
\subsection{Intermediate Coupling}
\begin{figure}
   \includegraphics[scale=0.65]{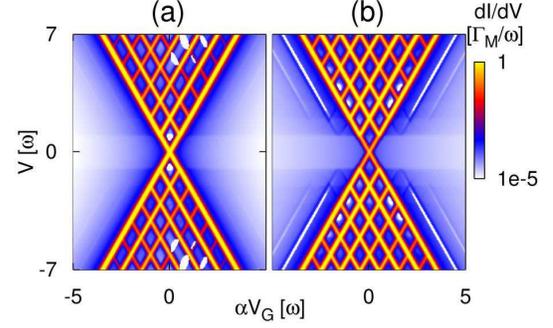}
   \caption{
     \label{fig:3}
   (Color online).
   Differential conductance as a function of gate and bias voltage close to the
   $N=0 \leftrightarrow N=1$ degeneracy point, for $\lambda = 1$ (a) and $\lambda = 2$ (b).
   In the logarithmic scale the lower end has to be chosen positive,
   preventing negative values from being correctly displayed, see e.g. white areas
   inside the sequential tunneling region in (a) which actually correspond to very weak 
   negative differential conductance (NDC). 
   }
\end{figure}

The differential conductance as a function of gate and bias voltage is shown in Fig.~\ref{fig:3}
in the case of intermediate electron-vibration coupling, $\lambda = 1$ in (a) and $\lambda = 2$ in (b).
The regimes where SET processes give the main contribution to the current
are triangle-shaped regions emanating from the point $V_\text{G}=0$,
where the energy for electron addition without changing the vibrational
quantum number lies between the electro-chemical potentials.
Due to the quantized nature of the vibration of the SMT, additional sharp peaks appear
in the differential conductance,
associated with a change of the vibrational quantum number.
Since in the SET-regime this is accompanied by a change in the charge,
the positions of these peaks depend linearly on the applied gate voltage.
At the $k$-th resonance line (counting from $V=0$)
a new set of transitions becomes energetically allowed, where the vibrational quantum
number changes by $k$ upon 
(dis)charging.
\par
Outside these two regimes, SET processes are suppressed by Coulomb blockade and one charge state is stable.
Here no features are seen in a plot corresponding to Fig.~\ref{fig:3} calculated to lowest non-vanishing order
(not shown, see e.g.~\cite{Nowack05,Koch04b}).
However, since we include all next-to-leading order processes, distinct features appear in this region,
which we now discuss.
When the bias voltage reaches the vibrational level spacing,
inelastic cotunneling processes exciting one vibrational quantum become energetically allowed.
Due to the harmonic spectrum, this makes every excited vibrational state for fixed $N$ accessible through
a sequence of such tunneling processes: the molecular vibration is driven out of equilibrium.
Each inelastic process involves the virtual occupation of an adjacent charge state
with an arbitrary vibrational excitation number.
The onset of inelastic cotunneling is seen as steps in the differential conductance,
whose positions are independent of the gate voltage since the process does not change the charge state of the SMT.
The magnitude of the steps however depend on the gate voltage since
the occupation of the virtual intermediate state is algebraically suppressed with the energy of this state.
Similarly, at $V = k\omega$ inelastic cotunneling processes exciting $k$ vibrational quanta become possible.
The corresponding 2nd and 3rd inelastic cotunneling steps are weakly seen for $\lambda = 2$, while,
for the tunneling coupling considered here, the suppression of the corresponding
Franck-Condon factors renders them invisible for $\lambda = 1$.
\par
A striking difference between Fig.~\ref{fig:3}(a) and (b) is the appearance of gate-dependent
lines inside the Coulomb blockade region in (b). The gate-dependence indicates that these lines are due to
processes changing the charge state of the SMT, but they cannot be due to SET processes starting
out from the vibrational ground state, since these are exponentially suppressed by Boltzmann factors
(energy conservation). They originate instead from SET processes starting out from an excited vibrational
state, which has previously been occupied by inelastic cotunneling processes.
This sequence of leading and next-to-leading order tunneling processes is called cotunneling-assisted 
sequential tunneling (COSET)
~\citep{Golovach04, Schleser05, Koch06, Luffe07, Aghassi08}, in the context of inelastic electron tunneling
spectroscopy (IETS) often referred to as phonon absorption
peaks, see Ref.~\citep{Galperin07} and references therein.
For even larger electron-vibration coupling
these features become more pronounced as discussed in the next section.
\subsection{Cross-over to Strong Coupling \label{sec:strong}}
\begin{figure}[t!]
     	\includegraphics[scale=0.65]{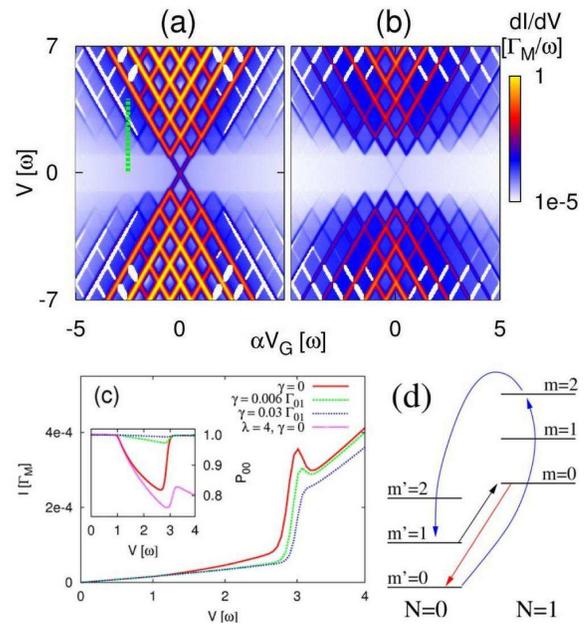}
     \caption{
     \label{fig:4}
   (Color online).
   (a) and (b): Differential conductance as a function of gate and bias voltage for strong
   electron-vibration coupling, $\lambda = 3$ in (a) and $\lambda = 4$ in (b). 
   For some values of the applied voltages the COSET processes result in
   closely spaced positive and negative differential conductance peaks, corresponding to peaks in the current.
   (c): Current as a function of bias voltage for $\lambda = 3$ along the dashed green line in (a), where
   the COSET processes give rise to a step + peak feature. As the vibrational relaxation rate, $\gamma$,
   is increased relative to $\Gamma_{01} = \Gamma |f_{01}|^2$ the peak vanishes, while the step remains.
   Inset: occupation of the vibrational ground state of the $N=0$ charge state, including also the result
   for $\lambda = 4$ without relaxation. 
   (d): Sketch of lowest vibrational states in the $N=0,1$ charge states. An example of a COSET process contributing
   to the step + peak in (c) consists of an inelastic cotunneling process (blue arrows), 
   followed by a sequential tunneling process (black arrow) into the
   vibrational ground state of the unstable charge state ($N = 1$). This may in turn sequentially relax (red arrow)
   to the vibrational ground state of the stable charge state ($N = 0$).
   }
\end{figure}

The results of the calculations for larger electron-vibration coupling are shown in Fig.~\ref{fig:4},
where $\lambda = 3$ in (a) and $\lambda = 4$ in (b).
The most obvious consequence of a large electron-vibration coupling is the suppression of the low bias
conductance. This Franck-Condon blockade stems from exponentially vanishing overlap integrals (Franck-Condon factors)
between low-lying vibrational states~\citep{Braig03a, Koch04b, Pistolesi07}
which is seen
in Fig.~\ref{fig:4} as a suppression of the degeneracy point peak (the differential conductance peak at
$V = V_\text{G} = 0$). In the case of an equilibrium vibrational distribution and lowest order transport
calculation, the current
would increase exponentially with increasing bias voltage, until the
blockade is lifted at around $V/2 = (\lambda^2 - \lambda) \omega = m$ corresponding to the first large
Franck-Condon factor $f_{m 0}$. However, when the vibrational distribution is pushed out
of equilibrium by sequential tunneling processes, 
this significantly enhances the current compared to the case of equilibrium vibrations. 
Additionally, when next-to-leading
order transport processes are taken into account, which was done
using the Golden-Rule T-matrix approach in Ref.~\citep{Koch06} to study the strong coupling regime 
($\lambda = 4$ and $\lambda = 5$), 
elastic and inelastic cotunneling processes change
the exponential suppression into an algebraic one.
Cotunneling processes take place through high lying virtual intermediate vibrational states ($m \sim \lambda^2$)
which have a large overlap with the vibrational ground state,
and the suppression of these processes is only algebraic
with respect to the energy of the virtual state, and therefore with respect to $\lambda$.
\par
The lowest inelastic cotunneling step is clearly seen for both $\lambda = 3$ and $\lambda =4$.
Additionally we find an anomalous signature of COSET processes.
For low bias voltage, just above the inelastic cotunneling threshold, these processes give rise to
positive differential conductance (PDC) features, i.e. current steps,
showing up as blue lines in Fig.~\ref{fig:4}.
However, at larger bias voltages for $\lambda = 3$ we observe pairs of white and blue lines,
corresponding to closely spaced lines of positive (PDC) and negative (NDC) differential conductance
(see note on logscale in caption of Fig.~\ref{fig:3}).
The nature of these line-pairs is more clearly seen in the current as a function of bias voltage, 
see red solid curve in Fig.~\ref{fig:4}(c): COSET gives rise to a step in the current
and, surprisingly, superimposed on it a peak.
Such a peak has not been reported previously to our knowledge and represents the central result of 
this section. We point out that 
all signatures in the current depend on a complicated interplay of a multitude of 
transport processes, also involving coherent superpositions of vibrational states. 
Basically, the peak arises due to a competition between leading and next-to-leading 
order transport processes
and is closely related to the non-equilibrium vibrational distribution.
This becomes clear from the inset of Fig.~\ref{fig:4}(c) where we show the occupation of the vibrational ground state
of the $N = 0$ charge state for bias voltages around the peak.
Although many vibrational excitations are involved, the sketch in Fig.~\ref{fig:4}(d) gives an indication of the types of relevant tunneling processes.
As the bias voltage exceeds the vibrational level spacing,
inelastic cotunneling (blue arrows in Fig.~\ref{fig:4}(d)) starts to deplete the ground state
in favor of higher laying vibrational states in the $N = 0$ charge state
(the first excited vibrational state  acquires almost all of the probability lost in the inset of (c)).
Cotunneling processes starting from the excited states now give a significant contribution to the current
which slowly increases with voltage.
As one approaches the threshold for COSET from below, the current sharply increases
as relaxation of these excited states into the $N=1$ states by sequential tunneling (black arrow) becomes energetically allowed
with large SET rates.
If the FC-blockade is not fully developed, a sequential tunneling process starting from $N=1$ into the $N=0$ ground state may now follow
with a larger rate than the inelastic cotunneling rate depleting the ground state, 
\emph{enhancing} its occupation.
As the voltage moves through the COSET resonance this feedback increasingly suppresses the contributions from
cotunneling processes starting from excited states, thereby suppressing the current.
As a result a thermally broadened peak occurs on top of the current step  in Fig.~\ref{fig:4}(c).
More generally, such peaks appear when cotunneling processes start to become significant 
($\lambda$ not too small) and compete with sequential tunneling processes, not fully suppressed by
Franck-Condon blockade ($\lambda$ not too large).  
\par
The effects of relaxation of the vibrational distribution due to
a coupling to a dissipative environment, i.e. to substrate phonon modes, now has an interesting effect:
as it is increased, at first it only suppresses the peak by disrupting the above competition. 
To illustrate this we have included a relaxation rate
on a phenomenological level
through an additional rate matrix ${\bf W}_\text{relax}$. This matrix is calculated in the same
way as the tunneling rate matrix ${\bf W}$, by performing an analogous perturbation
expansion in the coupling to the dissipative bath, $\gamma$, with the difference that the bath 
operators are Bosonic rather than Fermionic. However, we here restrict ourselves to the limit of weak coupling to the
bath, $\gamma \ll \Gamma$,
in which case we can stop this expansion at lowest non-vanishing order,
analogous to Ref.~\cite{Wegewijs05}, and incorporate the result in the
4th order electronic rate matrix ${\bf W}^{(4)}_{dd}$. We emphasize that such a simplified treatment
becomes invalid as $\gamma \sim \Gamma$ since this requires treating coupling to the electron
and phonon reservoirs on an equal footing.
The results for finite $\gamma$ is shown in the green dashed and blue dotted curves in Fig.~\ref{fig:4}(c).
The step only vanishes when $\gamma > \Gamma_{01} = \Gamma |f_{01}|^2$ (not shown), causing the first excited
vibrational state to always relax before being emptied by sequential tunneling. The peak on the other hand depends
on allowing several cotunneling processes to take place between relaxation events, and is thus much more sensitive
to the coupling to the bath,
thereby providing an accurate experimental probe of the strength of the dissipative coupling.
Additionally, since it only occurs within a range of $\lambda \sim 2-3$ it 
also reveals information concerning the strength of the electron-vibration coupling.
\par
For $\lambda = 4$, 
we find qualitatively similar results as presented in Ref.~\cite{Koch06}.
The degeneracy point is almost completely invisible due to the strong Franck-Condon blockade and
the COSET processes do not give rise to peaks, but rather to PDC lines at low bias. The NDC lines seen
at higher bias running perpendicular to the Coulomb diamond edges occur already in a lowest order calculation within the sequential tunneling
region~\citep{Koch04b}. These NDC lines are seen to continue into the Coulomb blockade region in our
next-to-leading order calculation and are of a different origin.
The absence of peaks at low bias is due to the fully developed Franck-Condon blockade, suppressing SET between 
vibrational ground states, thereby breaking the feedback mechanism which generates the peaks. 
The pink, fine dotted line in the inset of Fig.~\ref{fig:4}(c) shows the ground state occupation for 
$\lambda = 4, \gamma = 0$. It is clearly
seen that, in contrast to the $\lambda = 3$ case, the ground state does not become fully occupied 
above the threshold for COSET.
\section{Conclusions and outlook}\label{sec:Conclusions}
In this paper we have presented explicit kinetic equations for quantum transport, valid for a 
generic class of molecular quantum-dot type systems, accounting for all contributions up to 4th 
order perturbation theory in the tunneling Hamiltonian and the complete non-equilibrium molecular density matrix.
Due to the broadening of the states, which is treated correctly in the perturbation expansion, 
all terms are automatically well-defined for any set of system parameters.
The effective 4th order transition rates, coupling diagonal
elements of the molecular density matrix, include corrections from non-diagonal elements between \emph{non-degenerate states}.
In contrast to lowest order perturbation theory these corrections are essential 
for a physically correct solution.
Applying the theory to the specific model of a molecular transistor coupled to a localized vibrational mode,
we have shown that the signatures of cotunneling-assisted sequential tunneling become more pronounced as the
strength of the electron-vibration coupling is increased.  
In the cross-over to strong electron-vibration couplings, the
cotunneling-assisted SET processes were shown to give rise to
\emph{current peaks} in the Coulomb blockade regime, which signal a non-equilibrium vibrational state of
the molecule.
Their occurrence thus  provides an indication of strength of the electron-vibration interaction.
Since these peaks depend sensitively on an additional coupling to a dissipative bath,
they also provide a way to experimentally estimate this coupling strength, $\gamma$,
and thereby the important $Q$-factor ($Q = \omega / \gamma$).
\par
We acknowledge 
C. Emary for careful reading of the manuscript and 
many stimulating discussions with
 H. Schoeller, J. K\"onig, M. Hettler, J. Aghassi,
 F. Reckermann,
 K. Flensberg,
 J. Koch
and the financial support from
DFG SPP-1243,
the NanoSci-ERA, the Helmholtz Foundation and
the FZ-J\"ulich (IFMIT).
\appendix
\onecolumngrid
\section{Derivation of the kinetic equation\label{sec:kinetic_equation}}

\begin{figure}
   \includegraphics[scale=0.22]{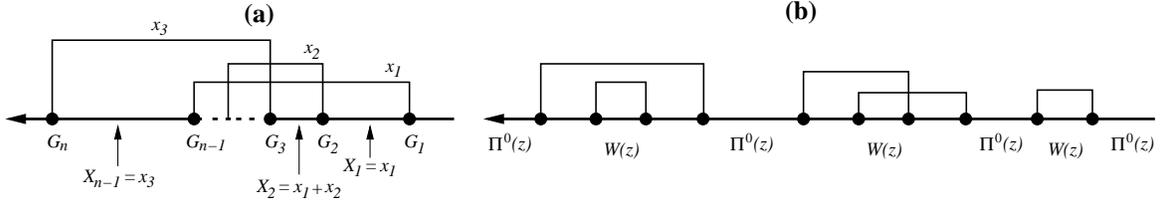}
   \caption{\label{fig:5}
     Diagrammatic representation of super-operator expressions. Processes
     evolve from right to left i.e. the diagrams have the same
     ordering as the expressions.
     (a): An example of an (irreducible) term in the expansion
     (\ref{eq:expansion}).
     (b): Separation into irreducible parts (self-energy or
     kernel, $W \left( z \right)$) and free evolution, $\Pi^0 \left( z \right)$.
     The rightmost diagram is the only one contributing to the leading order
     self-energy, $W^{\left( 2 \right)}$, while the two other diagrams are the only
     ones in next-to-leading order, contributing to $W^{\left( 4 \right)}$.
   }
\end{figure}

Our goal here is to derive the propagation of the reduced density matrix in Laplace space~(\ref{eq:laplace_rho3})
starting from Eq. (\ref{eq:laplace_rho2}). In the process we derive
all diagrammatic rules. 
A number of techniques exist for calculating the
trace over the reservoirs explicitly, such as projection operator 
techniques~\cite{Gardiner_noisebook}
or path integral methods~\cite{Legget87}.
Although being formally equivalent to a diagrammatic 
expansion on a Keldysh double contour, see e.g. Ref.~\cite{Schoeller94}, the diagram technique derived below has
a number of advantages: (i) it is completely formulated and derived in Laplace
space, (ii) a minimal number of diagrams represents all contributions in a
given perturbation order, Keldysh and electron/hole indices being summed over,
(iii) diagrams represent super-operators with diagram rules formally very
similar to those for operators. This means we can postpone taking matrix
elements, where the peculiarities of the Keldysh indices explicitly enter, 
to the end.
Expanding the denominator in (\ref{eq:laplace_rho2}) we have
\begin{eqnarray}
\label{eq:Pexpansion}
  P \left( z \right) = i \underset{\text{R}}{\text{Tr}} \left\{ \frac{1}{z -
  L_{\text{R}} - L} + \frac{1}{z - L_{\text{R}} - L}
  L_{\text{T}} \frac{1}{z - L_{\text{R}} - L} L_{\text{T}}
  \frac{1}{z - L_{\text{R}} - L} + \ldots . \right\} P \left( 0
  \right) \rho_{\text{R}}, 
\end{eqnarray}
where $\left( z - L_{\text{R}} - L \right)^{- 1}$ is the free
propagator and only even powers in $L_{\text{T}}$ give a non-vanishing
contribution when performing the trace. \ The crucial step in developing a
compact formalism is to ensure from the outset that Wick's theorem 
can be applied to \emph{super-operators} 
in the same way as for \emph{operators}.
This is achieved by the definition of dot (\emph{G}) and reservoir
(\emph{J}) super-operators by their action on an arbitrary operator
$A$:
\begin{eqnarray}
\label{eq:Gdef}
  G^p_{r \sigma \eta} A & = & p^{N_G} \sum_N \sum_{a_{1 p} \in N}^{a_{2 p} \in
  (N + p \eta)} T_{r \sigma \left( p \eta \right)}^{a_{2 p} a_{1 p}}  \left\{
  \begin{array}{lll}
    \phantom{- A}|a_{2 +} \rangle \langle a_{1 +} | A, &  & p = +\\
            -  A|a_{1 -} \rangle \langle a_{2 -} |, &  & p = -
  \end{array} \right. \label{eq:vertex} 
\end{eqnarray}
 \begin{eqnarray}
  J^p_{r \sigma \eta \omega} A & = &  \left\{ \begin{array}{lll}
    c_{r \sigma \eta \omega} A, &  & p = +\\
    A c_{r \sigma \eta \omega}, &  & p = -
  \end{array} \right. 
\end{eqnarray}
where we have assumed the tunneling matrix elements to be real-valued.
Here $p = \pm$ is a Keldysh index, distinguishing between the forward ($\left.
p = + \right)$ and backward ($\left. p = - \right)$ time-evolution on a
standard Keldysh double contour diagram. The index $\eta = \pm$ indicates an
annihilation / a creation reservoir field operator. The product $p \eta = \pm$
has a physical meaning: when acting with $G^p_{r \sigma \eta}$ on a density operator,
an electron / a
hole is added to the dot from electrode $r$ by projection between dot states
with different charge and spin. The amplitude involves the tunneling matrix
element and a Keldysh sign $p$. An additional Keldysh sign
$p^{N_G}$ appears in the amplitude. Importantly it can be assigned in any super-operator expression
(i.e. without taking matrix elements) by simply counting the number $N_G$ of
$G$s standing to the left (i.e. at later times). 
The explicit matrix elements
of $G$ (c.f. Eq. (\ref{eq:supermatrix-element})) required below are
\begin{eqnarray}
  \left. (G^p_{r \sigma \eta} \right)^{a_{1 +} a_{1 -}}_{a_{2 +} a_{2 -}} =
  p^{1 + N_{a_{2 +}} - N_{a_{2 -}}} \, \, \,T_{r \sigma \left( p \eta
  \right)}^{a_{2 p} a_{1 p}} \delta_{a_{2 \bar{p}} a_{1 \bar{p}}},
  \label{eq:Gmatel} 
\end{eqnarray}
where $\bar{p} = -p$.
Here the Keldysh sign is written as the parity of the charge difference
between the final state of the $G$ i.e. $\left( - 1 \right)^{N_{a_{2 +}} -
N_{a_{2 -}}} = \left( - 1 \right)^{N_G}$
(to see this, use that acting with $L_\text{T}$ ($\sim G$) changes the charge difference between the 
forward and backward contour of a Keldysh diagram by $\pm 1$, and that each diagram must 
start and end in a state which is diagonal in charge due to charge conservation of the 
total system).
With these definitions it can be verified that the
interaction $L_{\text{T}}$ can be written as
\begin{eqnarray}
\label{eq:tunnelingL}
  L_{\text{T}} = \sum_{p r \sigma \eta} \int d \omega p^{N_G} G^p_{r \sigma \eta}
  J^p_{r \sigma \eta \omega} \rightarrow p^{N_{G_i}}_i G^{p_i}_i J^{p_i}_i, 
\end{eqnarray}
where in the second form we have defined the short-hand indices $i = r_i
\sigma_i \eta_i \omega_i$ and implicitly sum over $p_i, r_i, \sigma_i, \eta_i$
and integrate over $\omega_i$. The reservoir super-operators satisfy
$L_{\text{R}} J^{p_i}_i = J^{p_i}_i \left( L_{\text{R}} - x_i \right)$ where
$x_i = \eta_i \omega_i$. In each term in the expansion,
\begin{equation}
  \text{$\begin{array}{lll}
    \underset{\text{R}}{\text{Tr}} \frac{1}{z - L_{\text{R}} - L}
    L_{\text{T}} \frac{1}{z - L_{\text{R}} - L} L_{\text{T}} \ldots
    L_{\text{T}} \frac{1}{z - L_{\text{R}} - L} L_{\text{T}}
    \frac{1}{z - L_{\text{R}} - L} P \left( 0 \right)
    \rho_{\text{R}} \\
    \\
    = p_n^{N_{G_n}} \ldots p_1^{N_{G_1}}
    \left( \underset{\text{R}}{\text{Tr}} J_n^{p_n} \ldots J_1^{p_1} \rho_{\text{R}} \right)
    \frac{1}{z + X_n - L} G^{p_n}_n \frac{1}{z + X_{n - 1} -
    L} G^{p_{n - 1}}_{n - 1} \ldots G^{p_2}_2 \frac{1}{z + X_1 -
    L} G^{p_1}_1 \frac{1}{z - L} P \left( 0 \right) 
    &  & 
  \end{array}$} \label{eq:expansion}
\end{equation}
we can then pull all $J$s through to the left when adding $X_i = x_{1} + x_{2} +
\ldots + x_i$ to $L_{\text{R}}$ in the free propagators. Using
$L_{\text{R}} \rho_{\text{R}} = 0$, $\rho_\text{R}$ can be pulled
through as well. Since the reservoirs are assumed to be non-interacting we can
now apply Wick's theorem to evaluate the trace over the super-operators $J$.
In doing so one generates a Keldysh sign which exactly cancels 
$p_n^{N_{G_n}} \ldots p_1^{N_{G_1}}$. 
This motivates including the canceling sign in the dot~(\ref{eq:vertex}) and tunneling Liouvillian~(\ref{eq:tunnelingL})
super-operators to keep the final diagram rules simple.
We contract pairs of reservoir super-operators, each contraction giving a factor
\begin{eqnarray}
  \gamma_{ji} & \equiv & 
  p_i   
  \langle J^{p_j}_j J_i^{p_i} \rangle_{\text{R}} = p_i
  \delta_{r_j r_i} \delta_{\sigma_j \sigma_i} \delta_{- \eta_j, \eta_i} \delta
  (\omega_j - \omega_i) f (p_i (x_i - \eta_i \mu_{r_i}) / T_{r_i}), 
  \label{eq:gammaji} 
\end{eqnarray}
where $f (x) = (e^x + 1)^{- 1}$ is the Fermi-function and $T_{r_i}$ is the
temperature of reservoir $r_i$ (from hereon we assume equal temperatures of all
reservoirs, $T_{r_i} \equiv T$).
The Wick's sign follows in the usual way as the sign of the permutation which disentangles 
the contractions. All the Keldysh signs arise 
because the regular Wick's theorem can only be applied after all operators have been put on 
on the same forward Keldysh contour (i.e use cyclic invariance of the trace), see~\cite{Schoeller08-rtrg}
for details.
Each \emph{super-operator} in the
expansion (\ref{eq:expansion}) can thus be represented diagrammatically as
usual by a directed free propagator line, $\left( z + X_i - L
\right)^{- 1}$, interrupted by vertices $G_i^{p_i}$ which are contracted in
pairs. A contraction of super-operators $G_j^{p_j}$ and $G_i^{p_i}$ with $j >
i$ is represented by an undirected line, see in Fig.~\ref{fig:5}(a). Since 
$\eta_j = - \eta_i, \omega_j = \omega_i, \sigma_j = \sigma_i, r_j = r_i$ 
are enforced by the contraction, it can
be unambiguously labeled by the indices of $x_i, \sigma_i, \eta_i, r_i$ of the
earliest vertex $G_i^{p_i}$. The sum in $X_i$ collects only those $x$ indices of
lines passing over the free propagator segment $i$ (contraction lines of one vertex to
the left and one to the right), the other ones cancel.
We now collect into $ W \left( z \right)$ all \emph{irreducible}
diagrams, i.e. those where any vertical cut will hit at least one contraction line,
and let $\Pi^0 \left( z \right) = i\left( z - L \right)^{- 1}$ be
the contributions from free evolution of the molecule, see Fig.~\ref{fig:5}(b). The
molecular density matrix in Laplace space is now given by
\begin{eqnarray}
  P \left( z \right) = \frac{i}{z - L} \sum_{n = 0}^{\infty} \left(
  W \left( z \right) \frac{i}{z - L} \right)^n P \left( 0 \right)
  = \frac{i}{z - L - i W \left( z \right)} P \left( 0 \right), 
  \label{eq:Pz} 
\end{eqnarray}
where in the last step we have arrived at Eq. (\ref{eq:laplace_rho3}). The
expectation value of the current operator ${\hat I}_r$ is calculated analogously:
\begin{eqnarray}
  \langle \hat{I}_r \rangle \left( z \right) = \text{Tr} {\hat I}_r \rho \left( z \right) =
  \text{Tr} L_{I_r} \frac{i}{z - L_{\text{R}} - L - L_{\text{T}}} P
  \left( 0 \right) \rho_{\text{R}} & = & \underset{\text{M}}{\text{Tr}} W_{I_r} \left( z
  \right) P \left( z \right). \label{eq:Iz} 
\end{eqnarray}
In contrast to Eq. (\ref{eq:laplace_rho2}) we trace over the full system,
molecule + reservoirs ($\text{Tr} = \text{Tr}_\text{R} \text{Tr}_\text{M}$). Under the trace the action of the current operator 
$\hat{I}_r$ on an arbitrary
operator $A$ has been expressed using the super-operator $L_{I_r} A =
\frac{1}{2} \left\{ \hat{I}_r, A \right\}$ (\emph{anti}-commutator) which takes the
same form as $L_{\text{T}}$:
\begin{eqnarray}
  L_{\text{} I_r} \rightarrow \left( G_{I_r} \right)^{p_i}_i J^{p_i}_i. 
\end{eqnarray}
Going through similar steps as above, we introduce a kernel $W_{I_r} \left( z
\right)$ which differs from $W \left( z \right)$ only by having the last $G$
vertex replaced by a current vertex $G_{I_r}$ with matrix elements:
\begin{eqnarray}
  \left( \left( G_{I_r} \right)^{p_i}_{r_i \sigma_i \eta_i} \right)^{a_{1 +}
  a_{1 -}}_{a_{2 +} a_{2 -}} = \delta_{\left( \eta_i p_i \right) +} \delta_{r
  r_i} \left( G^{p_i}_{r_i \sigma_i \eta_i}
  \right)^{a_{1 +} a_{1 -}}_{a_{2 +} a_{2 -}}. &  & \label{eq:GIr} 
\end{eqnarray}

\section{Diagrammatic rules and properties of the kernel
\label{sec:diagram_rules}}

The expression (\ref{eq:Pz}) is still formally exact, but requires summing up
\emph{all} irreducible diagrams to obtain the kernel, which in general is
not possible. We can write $W \left( z \right) = \sum_{k = 1}^{\infty}
W^{\left( 2 k \right)} \left( z \right)$, where $W^{\left( 2 k \right)} \left(
z \right)$ includes all terms with $2 k$ tunneling vertices, giving a
perturbative expansion in the tunneling Liouvillian $L_{\text{T}}$ i.e. $W^{\left( 2 k
\right)} \sim L_\text{T}^{2k}$. We now summarize the diagrammatic rules obtained in
Appendix \ref{sec:kinetic_equation} for calculating the zero-frequency $z = i 0$
contribution to the kernel:
\begin{eqnarray}
W^{\left( 2 k \right)} \left( i 0 \right) = - i \sum_{\text{contr.}}
   \left( \prod \gamma \right) (- 1)^{N_p} G^{p_{2 k}}_{2 k} \frac{1}{i 0 +
   X_{2 k - 1} - L} G^{p_{2 k - 1}}_{2 k - 1} \cdots G^{p_2}_2
   \frac{1}{i 0 + X_1 - L} G^{p_1}_1. \label{eq:W2k} 
\end{eqnarray}
Here one implicitly sums over all occurring Keldysh indices $p_i = \pm$ as well as
$r_i, \sigma_i, \eta_i$ and integrates over all occurring energies $x_i$.
\begin{enumerate}
  \item $\left( \prod \gamma \right)$: Draw $2 k$ vertices $G_i^{p_i}, i = 2
  k, \ldots, 1$ on a line. Connect pairs \ $G^{p_j}_j$, $G_i^{p_i}$ with $j >
  i$ by a line denoting a Wick's-contraction. Equate the indices of $G^{p_j}_j$
  to $r_i \sigma_i$, and $- \eta_i$ and multiply by
  \[ \gamma = p_i f (p_i (x_i - \eta_i \mu_{r_i}) / T). \]
  A vertex is contracted only to one other vertex and the contractions must be
  irreducible i.e. any vertical line through the diagram will cut at least one
  contraction line.
  
  \item $(- 1)^{N_p}$: Determine the Wick's-contraction sign by counting the
  number of crossings of tunneling lines in the diagram. The parity of this
  number equals the parity of $N_p$, the number of permutations required to
  disentangle the contractions.
  
  \item Assign a propagator $\left( i 0 + X_i - L \right)^{- 1}$ to
  segment $i$ between vertex operators $G^{p_{i + 1}}_{i + 1}$ and
  $G^{p_i}_i$. Here $X_i = \sum_{l = \text{conn}} x_l$ is the sum of the
  energies of contractions \emph{passing through} this segment i.e. the
  energies $x_l$ from all vertices $i > l$ on the right contracted with some
  vertex to the left of the segment.
  
  \item $\sum_{\text{contr.}}$: Perform 1-3 for every possible irreducible
  Wick's-contractions of the $2 k$ vertices and sum them up.
\end{enumerate}
The current kernel $W_{I_r}$ is obtained by the same rules with the exception
that the last vertex is replaced by the current vertex, $G_{2 k}^{p_{2 k}}
\rightarrow \left( G_{I_r} \right)_{2 k}^{p_{2 k}}$. Due to the additional
$\delta$-functions in the $G_{I_r}$ vertex (\ref{eq:GIr}), we need only
include terms where an electron is added to the molecule from reservoir $r$ in
the final vertex $\left( \eta p = + \right)$. Additionally, due to the trace
in Eq. (\ref{eq:Iz}) we only need matrix elements which are diagonal in final
states.

One can check that these rules exactly reformulate the rules for the
diagrammatic expansion of the kernels $W$ and $W_{I_r}$ formulated previously
{\cite{Schoeller94}}, but in a compact manner well suited for 
constructing the general transport rates considered here. Fig. \ref{fig:5}(b)
shows the \emph{single} diagram for the leading order $W^{\left( 2
\right)}$ and the \emph{two} diagrams making up the next-to-leading order
kernel $W^{\left( 4 \right)}$. Since a Liouville diagram of order $2 k$ has $2
k$ Keldysh indices $p$, as well as $k$ electron/hole indices $\eta$, these diagrams
account for $2_p^2 \times 2_{\eta}^{} = 8$ ($2 \times 2_p^4 \times 2_{\eta}^2 =
128$) different Keldysh diagrams in leading (next-to-leading) order! The
Keldysh representation is still useful in visualizing the character of the
involved tunneling processes but need not be considered here.

The general property of the kernel
$
\left( W (z) \right)^{a_{1+} a_{1-}}_{a_{2+} a_{2-}} = 
\left( W (- z^{\ast}) \right)^{a_{1-} a_{1+}\ast}_{a_{2-} a_{2+}}
$
guarantees a Hermitian stationary-state density matrix. In
the stationary limit ($z \rightarrow i 0$), we have
\begin{eqnarray}
\label{eq:WhermitianRe}
\text{Re} \left( W (i 0) \right)^{a_{1 +} a_{1 -}}_{a_{2 +} a_{2 -}} & =
& \phantom{-} \text{Re} \left( W (i 0) \right)^{a_{1 -} a_{1 +}}_{a_{2 -}
a_{2 +}},\\
\label{eq:WhermitianIm}
\text{Im} \left( W (i 0) \right)^{a_{1 +} a_{1 -}}_{a_{2 +} a_{2 -}} & =
& - \text{Im} \left( W (i 0) \right)^{a_{1 -} a_{1 +}}_{a_{2 -} a_{2 +}}.
\end{eqnarray}
This has the important implication that elements of the kernel which are
diagonal in the double-indices $a_{1 +} = a_{1 -}$ and $a_{2 +} = a_{2 -}$ are
real-valued since they contain pairs of diagrams represented by complex conjugate
expressions (obtained by inverting all $p$ and $\eta$ indices on a diagram).
The same holds for $W_{I_r} \left( i 0 \right)$.

Additionally, the charge difference between forward and backward Keldysh contours 
is conserved by each diagram. 
To see this, consider the action of a vertex operator $G^{p_i}_{r_i \sigma_i \eta_i}$,
which changes the charge number on contour $p_i$ by $p_i \eta_i$.
Since in Eq.~(\ref{eq:W2k}) this is contracted with $G^{p_j}_{r_j \sigma_i -\eta_i}$, 
this change in charge is either canceled ($p_j = p_i$) or equals that on the opposite contour 
($p_j = -p_i$). The same hold for all pairs of contractions.

\section{2nd order
\label{sec:2ndorder}}

In 2nd order, there is only one Liouville diagram, see Fig.
\ref{fig:5}(b). The diagrammatic rules give (we omit the argument $i 0$)
\begin{eqnarray}
  W^{(2)}
  & = &
  - i \gamma_{21} G^{p_2}_2 \frac{1}{i 0 + x_1 - L} G^{p_1}_1. 
\end{eqnarray}
The explicit evaluation of the matrix elements of this expression is discussed
in some detail now, so that it can be skipped in the 4th order calculation where the
expressions are less transparent, obscuring the basic simple operations. We
introduce a shorthand notation for states on the forward / backward
propagators: $a_i \equiv a_{i +} a_{i -}$ and their energy difference $E_{a_i}
\equiv E_{a_{i +}} - E_{a_{i -}}$. Taking matrix elements and explicitly
writing out summations and integrations, we obtain
\begin{eqnarray}
  \label{eq:W2superform}
     \left( W^{(2)} \right)^{a_0}_{a_2}
     & = &
     - i \sum_{p_2 p_1} \sum_{r_1 \sigma_1 \eta_1} \sum_{a_{1 \pm}}
     \left( G^{p_2}_{\bar{\eta}_1 r_1 \sigma_1} \right)^{a_1}_{a_2}
     \left( G^{p_1}_{\eta_1 r_1 \sigma_1} \right)^{a_0}_{a_1}
     \int dx_1 \frac{ p_1 f (p_1 (x_1 - \eta_1 \mu_{r_1}) / T) }
                    { i 0 + x_1 - E_{a_1} }
     \\
     & = &
     - i \sum_{p_2 p_1} \sum_{r_1 \eta_1} \sum_{a_{1 \pm}}
     {  p_2 p_1 } \times
     \left( \sum_{\sigma_1}
       T_{r_1 \sigma_1 \left( \bar{\eta}_1 p_2 \right)}^{a_{2 p_2} a_{1 p_2}}
       T_{r_1 \sigma_1 \left(       \eta_1 p_1 \right)}^{a_{1 p_1} a_{0 p_1}}
     \right)
     \delta_{a_{2 \bar{p}_2} a_{1 \bar{p}_2}} \delta_{a_{1 \bar{p}_1} a_{0 \bar{p}_1}} \nonumber
     \\
  \label{eq:W2}
     && \times
     \left( -{ p_1} \phi ( \left( E_{a_1} - \eta_1 \mu_{r_1} \right) / T)
           - i \pi f (p_1 \left( E_{a_1} - \eta_1 \mu_{r_1}) / T \right) \right),
\end{eqnarray}
where $\bar{\eta}_1 = - \eta_1$ and $\bar{p}_1 = - p_1$. The overall sign $p_1
p_2$ arises from several contributions. There is no Wick's sign since there is
only one contraction (rule 2). The contraction-function gives a sign $p_1$.
Finally, the matrix elements of the vertices involve a sign $p_2 p_1$ and
additionally a sign $p_1$ since $G_1^{p_1}$ has an odd number of $G$s standing
to its left.

For the integration we assume a flat density of states with a large bandwidth
$D \gg T, E_{a_1} - \mu_r, \mu_r - \mu_{r'}$ i.e. all energies $E_{a_1}$ lie
deep within this band, including all $\mu_r$, 
meaning that we can neglect terms
proportional to $\int_{D - V}^{D} d x \frac{1}{x} \approx  V / D \ll 1$. 
Using $\frac{1}{x + i 0} = P \frac{1}{x} - i \pi
\delta \left( x \right)$, where $P$ denotes the principal value, we split up the integral into real and imaginary
parts. The imaginary part involves the Fermi-function and is the only contribution to 
elements of $W$ diagonal in initial and final indices, which are just the well-known
Golden Rule rates. The real part is only relevant for elements of $W$ which
are off-diagonal in initial or final indices, and involves the function ($x$
rescaled by $T$)
\begin{eqnarray}
  \phi (\lambda)
  & = &
  - \text{Re} \int_{- \frac{D}{T}}^{\frac{D}{T}} dx
  \frac{f(x)}{i 0 + x - \lambda}
  \nonumber \\
  & = & 
  - \text{Re} \psi \left( \frac{1}{2} + i \frac{\lambda}{2 \pi} \right) + \ln \frac{D}{2 \pi T},
  \label{eq:phi}
\end{eqnarray}
where $\lambda = \left( E_{a_1} - \eta_1 \mu_{r_1} \right) \text{/} T$ and
$\psi$ is the digamma function. To arrive at this form we have used \ $f (px)
= (1 - p) / 2 + pf (x)$ and neglected the integral $\text{Re} \int_{- D/T}^{D/T} dx
\frac{(1 - p) / 2}{i 0 + x - \lambda} \propto \lambda T / D$. Clearly, $\phi
\left( \lambda \right)$ is symmetric for real-valued arguments, and we may
write $\phi ( \left( E_{a_1} - \eta_1 \mu_{r_1} \right) / T) = \phi \left(
\left( \eta_1 E_{a_1} - \mu_{r_1} \right) / T \right)$ i.e. only the distance
of the addition energy to the Fermi-energy is relevant, irrespective of
whether it is an electron / hole process ($p_1 \eta_1 = \pm$).
The curve has a peak 
$\phi(0) = \gamma_\text{E} + 2 \text{ln} 2 + \text{ln} \frac{D}{2\pi T} = 1.96351 + \text{ln} \frac{D}{2\pi T}$,
where $\gamma_\text{E}$ is the Euler's constant, 
and logarithmic tails, 
$\phi(\lambda) \approx \text{ln} \frac{D}{\lambda T}$ for $\lambda \gg 1$. 

\section{4th order
\label{sec:4thorder}}

In 4th order we have two irreducible contractions of the four vertices.
We refer to the first diagram (leftmost in Fig.~\ref{fig:5}(b)) as direct (D) type
and the second one (middle in Fig.~\ref{fig:5}(b)), which gets an additional sign 
from the Wick's contraction, as exchange (X) type.
Applying the diagrammatic rules we obtain
\begin{eqnarray}
  \label{eq:W4}
     W^{\left( 4 \right)} (i 0)
     & = &
     - i \gamma_{32} \gamma_{41}
     G^{p_4}_4 \frac{1}{i 0 + x_1       - L}
     G^{p_3}_3 \frac{1}{i 0 + x_1 + x_2 - L}
     G^{p_2}_2 \frac{1}{i 0 + x_1       - L}
     G^{p_1}_1
     \hspace{0.5cm}
     \text{(D)}
     \\
     &  &
     + i \gamma_{42} \gamma_{31}
     G^{p_4}_4 \frac{1}{i 0 + x_2       - L}
     G^{p_3}_3 \frac{1}{i 0 + x_1 + x_2 - L}
     G^{p_2}_2 \frac{1}{i 0 + x_1       - L}
     G^{p_1}_1.
     \hspace{0.5cm}
     \text{(X)}
\end{eqnarray}
Taking matrix elements, expanding all indices and
explicitly writing out all summations and integrations this becomes 
\begin{eqnarray}
  \label{eq:W4trans}
    &  & \left( W^{(4)} \right)^{a_0}_{a_4} = - i
    \sum_{p_4 p_3 p_2 p_1} \sum_{r_2 r_1} \sum_{\sigma_2 \sigma_1} \sum_{\eta_2 \eta_1} \sum_{a_{3 \pm} a_{2 \pm} a_{1 \pm}}
    \nonumber
    \\
    \nonumber
    \\
    &  &
    \left[
    \left( G^{p_4}_{\bar{\eta}_1 r_1 \sigma_1} \right)^{a_3}_{a_4}
    \left( G^{p_3}_{\bar{\eta}_2 r_2 \sigma_2} \right)^{a_2}_{a_3}
    \left( G^{p_2}_{\eta_2 r_2 \sigma_2} \right)^{a_1}_{a_2}
    \left( G^{p_1}_{\eta_1 r_1 \sigma_1} \right)^{a_0}_{a_1}
    \iint  dx_1 dx_2
    \frac{ p_2 p_1 f (p_2 (x_2 - \eta_2 \mu_{r_2}) / T)
                   f (p_1 (x_1 - \eta_1 \mu_{r_1}) / T)}
         { (i 0 + x_1 - E_{a_3}) (i 0 + x_1 + x_2 - E_{a_2}) (i 0 + x_1 - E_{a_1}) } \right.
    \nonumber
    \\
    \nonumber
    \\
    &  &
    \left.
    -
    \left( G^{p_4}_{\bar{\eta}_2 r_2 \sigma_2} \right)^{a_3}_{a_4}
    \left( G^{p_3}_{\bar{\eta}_1 r_1 \sigma_1} \right)^{a_2}_{a_3}
    \left( G^{p_2}_{\eta_2 r_2 \sigma_2} \right)^{a_1}_{a_2}
    \left( G^{p_1}_{\eta_1 r_1 \sigma_1} \right)^{a_0}_{a_1}
    \iint   dx_1 dx_2
    \frac{p_2 p_1 f (p_2 (x_2 - \eta_2 \mu_{r_2}) /  T)
                  f (p_1 (x_1 - \eta_1 \mu_{r_1}) / T)}
          { (i 0 + x_2 - E_{a_3}) (i 0 + x_1 + x_2 - E_{a_2}) (i 0 + x_1 - E_{a_1}) } \right].
    \nonumber
    \\
\end{eqnarray}
Note that the two expressions differ only by the lower indices of vertex 3 and
4 and by the electron frequency $x_1, x_2$ in the propagator connecting these
vertices.

For a non-degenerate spectrum, as discussed in the main text, we need only the
expressions for $W^{(4)}$ with $a_{4 +} = a_{4 -}$ and $a_{0 +} = a_{0 -}$,
which are guaranteed to be real-valued. We first give the
final, explicit result, before discussing how to arrive there.
\begin{eqnarray}
\label{eq:W4explicit}
    & &
    \text{Re} \left( W^{(4)} \right)^{a_{0 +} a_{0 -}}_{a_{4 +} a_{4 -}} =
    \frac{1}{T}
    \sum_{p_4 p_3 p_2 p_1} \sum_{r_2 r_1} \sum_{\eta_2 \eta_1} \sum_{a_{3 \pm} a_{2 \pm} a_{1 \pm}}
    \delta_{a_{4 \bar{p}_4} a_{3 \bar{p}_4}}
    \delta_{a_{3 \bar{p}_3} a_{2 \bar{p}_3}}
    \delta_{a_{2 \bar{p}_2} a_{1 \bar{p}_2}}
    \delta_{a_{1 \bar{p}_1} a_{0 \bar{p}_1}} p_4 p_1 
    \nonumber \\
    & &
    \phantom{-} \left\{ \left( 
      \sum_{\sigma_2} T_{r_2 \sigma_2 (\bar{\eta}_2  p_3)}^{a_{3 p_3} a_{2 p_3}} T_{r_2 \sigma_2 (\eta_2 p_2)}^{a_{2 p_2} a_{1p_2}}
      \sum_{\sigma_1} T_{r_1 \sigma_1 (\bar{\eta}_1 p_4)}^{a_{4 p_4} a_{3 p_4}}  T_{r_1 \sigma_1 (\eta_1 p_1)}^{a_{1 p_1} a_{0 p_1}}
    \right) \right.
    \nonumber \\
    & &
     \left[ p_2 p_1
      \frac{F (\left(E_{a_2} - \eta_1 \mu_{r_1} - \eta_2 \mu_{r_2}\right)/T, \left(E_{a_3} - \eta_1 \mu_{r_1}\right)/T)
          - F (\left(E_{a_2} - \eta_1 \mu_{r_1} - \eta_2 \mu_{r_2}\right)/T, \left(E_{a_1} - \eta_1 \mu_{r_1}\right)/T)}
           {\left(E_{a_3} - E_{a_1}\right)/T}
     \right.
    \nonumber \\
    & &
    \left.
         + p_1 (1 - p_2)
      \frac{\tilde{F} (\left(E_{a_3} - \eta_1 \mu_{r_1}\right)/T)
          - \tilde{F} (\left(E_{a_1} - \eta_1 \mu_{r_1}\right)/T)}
           {\left(E_{a_3} - E_{a_1}\right)/T} \right]
    \nonumber \\
    & &
    -
    \left(
      \sum_{\sigma_2} T_{r_2 \sigma_2 (\bar{\eta}_2 p_3)}^{a_{4 p_4} a_{3 p_4}} T_{r_2 \sigma_2 (\eta_2 p_2)}^{a_{2 p_2} a_{1 p_2}}
      \sum_{\sigma_1} T_{r_1 \sigma_1 (\bar{\eta}_1 p_4)}^{a_{3 p_3} a_{2 p_3}} T_{r_1 \sigma_1 (\eta_1 p_1)}^{a_{1 p_1} a_{0 p_1}}
    \right)
    \nonumber p_2 p_1 \\
    & &
    \left[
      \frac{ F (\left(E_{a_2} - \eta_1 \mu_{r_1} - \eta_2 \mu_{r_2}\right)/T, \left(E_{a_1} - \eta_1 \mu_{r_1}\right)/T)
      - F (\left(E_{a_3} + E_{a_1} - \eta_1 \mu_{r_1} - \eta_2 \mu_{r_2}\right)/T, \left(E_{a_1} - \eta_1 \mu_{r_1}\right)/T)}
           {\left(E_{a_2} - E_{a_3} - E_{a_1}\right)/T}
    \right.
    \nonumber \\
    & &
    \left. \left.
    +
    \frac{ F (\left(E_{a_2} - \eta_1 \mu_{r_1} - \eta_2 \mu_{r_2}\right)/T, \left(E_{a_3} - \eta_2 \mu_{r_2}\right)/T)
    - F (\left(E_{a_3} + E_{a_1} - \eta_1 \mu_{r_1} - \eta_2 \mu_{r_2}\right)/T, \left( E_{a_3} - \eta_2 \mu_{r_2}\right)/T)}
         {\left(E_{a_2} - E_{a_3} - E_{a_1}\right)/T} \right] \right\},
    \nonumber \\
\end{eqnarray}
where only two types of functions enter
\begin{eqnarray}
     F \left( \lambda', \lambda \right)
     & = &
     \pi \left\{
         \phi (\lambda' - \lambda) f (\lambda)
       + b (\lambda') \left[ \phi (\lambda' - \lambda) - \phi (- \lambda) \right]
     \right\}
      \label{eq:F}\\
     & \rightarrow &
      \pi \left\{
        \phi (- \lambda) f (\lambda)
        - \frac{d}{d \lambda} \phi (- \lambda)
      \right\}
      \text{ for $\lambda' \rightarrow 0$},
     \nonumber \\
      \tilde{F} \left( \lambda \right)
      & = &
       \frac{\pi}{2} \phi (\lambda),
      \label{eq:Ftilde}
\end{eqnarray}
where
$f ( \lambda ) = (e^{\lambda} + 1)^{- 1}$
and
$b ( \lambda ) = ( e^{\lambda} - 1 )^{- 1}$
are the Fermi- and Bose-function respectively and
$\phi (\lambda)$ is given by Eq.~(\ref{eq:phi}).
All expressions arising
from the integrals are explicitly seen to be well behaved,
since they take the form of differential quotients:
whenever a denominator vanishes, the numerator also vanishes with the same power,
resulting in a finite value.
The rates are thus well-behaved functions of all model parameters including the voltages.

We now discuss the steps leading from Eq. (\ref{eq:W4trans}) to Eq.
(\ref{eq:W4explicit}). The tunnel matrix elements enter automatically via the
vertices Eq. (\ref{eq:Gmatel}). The four vertices give a sign $p_4 p_3 p_2
p_1$, and the vertices $G^{p_3}_3$ and $G_1^{p_1}$ give an additional sign
$p_3 p_1$ (since they are followed by an odd number of vertices towards the
left). Combined with the contraction signs $p_2 p_1$ we get in total a sign $p_4 p_1$
for both diagrams.

The remaining task is to obtain the closed-form expressions for the imaginary part of the 
two integrals. Normalizing the integration variables to $T$ and then shifting them
introduces the energy denominators $\lambda_1 = \left( E_{a_1} - \eta_1
\mu_{r_1} \right) \text{/} T$ and $\lambda_2 = \left( E_{a_2} - \eta_1
\mu_{r_1} - \eta_2 \mu_{r_2} \right) \text{/} T$. For the last propagator we
get $\lambda_3 = \left( E_{a_3} - \eta_1 \mu_{r_1} \right) \text{/} T$ for the
$D$ type and $\lambda_3 = \left( E_{a_3} - \eta_2 \mu_{r_2} \right) \text{/}T$ 
for the $X$ type diagram. The integrals are then split into
partial-fractions:
\begin{eqnarray}
     I_D^{p_2 p_1}
     & = &
     \frac{1}{T} \iint   dx_1 dx_2
     \frac{f (p_2 x_2) f (p_1 x_1) }{ \lambda_3 - \lambda_1}
     \text{Im} 
     \left( \frac{1}{i 0 + x_1 + x_2 - \lambda_2} \right)
     \left(
         \frac{1}{i 0 + x_1 - \lambda_3}
       - \frac{1}{i 0 + x_1 - \lambda_1}
     \right),
     \\
     I_X^{p_2 p_1}
     & = &
     \frac{1}{T} \iint   dx_1 dx_2
     \frac{f (p_2 x_2) f (p_1 x_1)}{\lambda_2 - \lambda_3 - \lambda_1}
     \text{Im} 
     \left(
         \frac{1}{i 0 + x_1 + x_2 - \lambda_2}
       - \frac{1}{i 0 + x_1 + x_2 - \lambda_3 - \lambda_1}
     \right)
     \nonumber
     \\
     & & 
     \times
     \left(
         \frac{1}{i 0 + x_1 - \lambda_1}
       + \frac{1}{i 0 + x_2 - \lambda_3}
     \right),
\end{eqnarray}
where $I_D$ denotes the integral in~(\ref{eq:W4trans}) in the $D$ type 
and $I_X$ the one in the $X$ type diagram.
These can be expressed in the integrals encountered in 2nd order.
This is done most efficiently by first noting a number of \emph{sumrules}
which are satisfied by the integrals (but not by the diagrams!)
in the wide-band limit:
\begin{eqnarray}
  \sum_{p_1 = \pm} I_D^{p_2 p_1} =
  \sum_{p_1 = \pm} I_X^{p_2 p_1} =
  \sum_{p_2 = \pm} I_X^{p_2 p_1} = 0.
\end{eqnarray}
Summing the integrals over a Keldysh index $p_i$ we eliminate one Fermi-function using
$\sum_{p_i = \pm} f \left( p_i x_i \right) = 1$. We can then first evaluate
the integral over $x_i$ on the same contour as for the 2nd order integral, see
Appendix \ref{sec:contraction_integral}. If the integrand vanishes faster than $x_i^{- 1}$ the
contribution can be neglected in the wide-band limit, even when performing also
the 2nd integral. From the original expressions for the integrals in Eq.
(\ref{eq:W4trans}) one sees that this is the case, except for the integrand $I_D$
considered as function of $x_2$.
Therefore $\sum_{p_2 = \pm} I_D^{p_2 p_1} \neq 0$.
This implies that $I_X$ is proportional to $p_1 p_2$, while $I_D$ additionally
contains a term proportional only to $p_1$:
\begin{eqnarray}
     I^{p_2 p_1}_D
     & = &
     \frac{1}{T} p_2 p_1
     \frac{F (\lambda_2, \lambda_3) -  F (\lambda_2, \lambda_1)}
          {\lambda_3 - \lambda_1}
     + \frac{1}{T} p_1 (1 - p_2)
     \frac{\tilde{F} (\lambda_3) - \tilde{F} (\lambda_1)}
          {\lambda_3 - \lambda_1},
     \\
     I^{p_2 p_1}_X
     & = &
     \frac{1}{T} p_2 p_1
     \frac{  F (\lambda_2, \lambda_1) - F (\lambda_3 + \lambda_1, \lambda_1)   
           + F (\lambda_2, \lambda_3) - F(\lambda_3 + \lambda_1, \lambda_3)}
          {\lambda_2 - \lambda_3 - \lambda_1}.
\end{eqnarray}
It remains to be shown that $F (\lambda', \lambda)$ and $\tilde{F} \left(
\lambda \right)$ actually are given by Eq.~(\ref{eq:F}) and (\ref{eq:Ftilde}) respectively.
To do this we now use the expansion
\begin{eqnarray}
  f \left( p_2 x_2 \right) f \left( p_1 x_1 \right)
  &=&
  p_2 p_1 f\left(x_1\right) f\left(x_2\right)
  + \frac{1}{2} p_1 \left(1 - p_2\right) f\left(x_1\right) + \frac{1}{2} p_2 \left(1 - p_1\right) f\left(x_2\right)
  + \text{const.}
\end{eqnarray}
As was noted when deriving the above sumrule, only terms containing the product 
$f\left(x_1\right) f\left(x_2\right)$ give a non-vanishing contribution to the X-type integral, 
and we only have to consider integrals of the form
\begin{eqnarray}
  F \left( \lambda', \lambda \right)
  & \equiv  &
  \text{Im} \iint   dxdx'  \frac{f \left( x' \right) f \left( x \right)}{\left( i 0 + x + x'
      - \lambda' \right) \left( i 0 + x - \lambda \right)}
  \nonumber
 \\
     & = &
     - \pi \text{Re}
     \left[
      \int d x'  \frac{f (x') }{i 0 + x' - \lambda' + \lambda} f (\lambda)
       + b (\lambda')
       \int dx
       \frac{f (- x) - f (\lambda' - x)}{i 0 + x - \lambda}
  \right]
  \nonumber
  \\
     & = &
     \pi \left\{
         \phi (\lambda' - \lambda) f (\lambda)
         + b (\lambda') \left[ \phi (\lambda' - \lambda) - \phi (- \lambda) \right]
       \right\}.
\end{eqnarray}
Here we expanded $\text{Im} \left( \left( x + i 0
\right) \left( y + i 0 \right) \right)^{- 1} = - \pi \text{Re} \left\{ \delta
\left( x \right) \left( y + i 0 \right)^{- 1} + \delta \left( y \right) \left(
x + i 0 \right)^{- 1} \right\}$ and used the relation $f \left( x' \right) f
\left( x \right) = \left( f \left( - x' \right) - f \left( x \right) \right) b
\left( x + x' \right)$ in the second term.

The D-type integral gives a non-vanishing contribution also for terms containing 
only $f\left(x_1\right)$. This yields the additional integral where
$f \left( p_2 x_2 \right) f \left( p_1 x_1 \right) \rightarrow 
\frac{1}{2} p_1 \left( 1 - p_2 \right) f \left( x_1 \right)$
\begin{eqnarray}
     \tilde{F} \left( \lambda \right)
     & \equiv &
     \frac{1}{2} \text{Im} \iint   dx dx'
     \frac{ f \left( x \right)}{\left( i 0 + x + x' - \lambda' \right) \left( i 0 + x -
     \lambda \right)}
     \nonumber \\
     & = &
     - \pi \frac{1}{2} \text{Re} \int dx \frac{f (x)}{i 0 + x - \lambda}
     =  \frac{\pi}{2} \phi \left( \lambda \right).
 \end{eqnarray}
Note that $\lambda'$ drops out of the answer since $\text{Re} \int_{- D/T}^{D/T} dx'
\frac{1}{i 0 + \lambda + x' - \lambda'}$ vanishes for $D / T \gg \lambda,
\lambda'$.
\par
\section{Contraction integral
  \label{sec:contraction_integral}}
We comment on the calculation of the integral: 
\begin{eqnarray}
  \int d x \frac{f(x)}{i0 + x - \lambda}
  =
  + \text{Re} \psi \left( \frac{1}{2} + i \frac{\lambda}{2 \pi} \right)
  - \ln \frac{D}{2 \pi T} - i \pi f \left( \lambda \right).
\end{eqnarray}
It can be calculated with a smooth  Lorentzian cutoff of width
$D/T$ and the result must then be expanded in the small parameter $\lambda T / D$ to lowest order~\citep{Koenig_master}.
This however involves unnecessary complications since the energy scale separation
$D /T \gg \lambda$ is only used at the end.
Here we indicate how this may be avoided, simplifying this and other
similar calculations.
We first note that although 
$- i \pi f \left( \lambda \right)$ clearly stems from
$ i \text{Im}  \frac{1}{z - \lambda + i 0} = - i\pi \delta \left( z - \lambda
\right)$ one should not separate real and imaginary parts until the end
of the calculation.
We apply the residue theorem for a contour along the real axis and
\emph{finite} semi-circle in the upper half-plane i.e. not containing
the pole $z=\lambda-i0$. We obtain the integral with a sharp cutoff:
\begin{eqnarray}
     \int_{- D/T}^{D/T} d x \frac{f(x)}{x - \lambda + i 0}
     & = &
     - i \left. \int_0^{\pi} d \varphi  z
     \frac{f(z)}{z - \lambda + i 0} \right|_{z =  D e^{i \varphi} / T}
     - \left. 2 \pi i  \sum_{k = 0}^{k_D}  \frac{1}{z -
         \lambda + i 0} \right|_{z = i \pi \left( 1 + 2 k \right)},
\end{eqnarray}
where $k_D = \left[ \frac{D}{2 \pi T} - \frac{1}{2} \right]$ 
($\left[ \bullet \right]$ denotes the integer part).
We now explicitly calculate the contribution to the contour accounting for $D / T \gg \lambda$.
The latter is trivial since $f \left( z \right)$ is equal to $1$
for $\frac{\pi}{2} < \arg z < \pi$ and 0 elsewhere for $z$ on a
semi-circle of radius $D / T \gg \lambda$, as one easily verifies.
Since the remaining part of the integrand is independent of $\arg
z$ on this contour, we get a contribution $-i { \frac{\pi}{2}}$.
In the limit $D/T \gg \lambda$ the summation over Matsubara-poles can be extended 
to infinity and 
gives a digamma function plus a term depending logarithmically on the band-width: 
\begin{eqnarray}
       &  & \sum_{k = 0}^{k_D}   \left( \frac{1}{k + \frac{1}{2}
       + i \frac{\lambda}{2 \pi}} - \frac{1}{k + 1} \right)
       + \gamma_\text{E} + \ln k_D
       \approx  - \psi \left( \frac{1}{2} + i \frac{\lambda}{2 \pi} \right)
       + \ln \frac{D}{2 \pi T}
\nonumber\\
       &=&  - \text{Re} \psi \left( \frac{1}{2} + i \frac{\lambda}{2 \pi} \right)
       + \ln \frac{D}{2 \pi T} + i \pi f \left( \lambda \right) -
       { i \frac{\pi}{2}},
\end{eqnarray}
where we added and subtracted the Euler's constant, $\gamma_\text{E} = \lim_{n\rightarrow \infty} \sum_{k=1}^{n+1}
1/k-\ln n$. The contribution from the arc cancels part of the
imaginary part $\text{Im} \psi(1/2+i x)  =  \pi \tanh (\pi x) / 2 = \pi (1/2 - f(2\pi x))$.
In contrast, if one takes a cutoff function to make the integral vanish along the
semi-circle for infinite radius~\cite{Koenig_master}, one unnecessarily complicates
the evaluation of the residues.
\twocolumngrid
\bibliographystyle{apsrev}

\begin{thebibliography}{48}
\expandafter\ifx\csname natexlab\endcsname\relax\def\natexlab#1{#1}\fi
\expandafter\ifx\csname bibnamefont\endcsname\relax
  \def\bibnamefont#1{#1}\fi
\expandafter\ifx\csname bibfnamefont\endcsname\relax
  \def\bibfnamefont#1{#1}\fi
\expandafter\ifx\csname citenamefont\endcsname\relax
  \def\citenamefont#1{#1}\fi
\expandafter\ifx\csname url\endcsname\relax
  \def\url#1{\texttt{#1}}\fi
\expandafter\ifx\csname urlprefix\endcsname\relax\def\urlprefix{URL }\fi
\providecommand{\bibinfo}[2]{#2}
\providecommand{\eprint}[2][]{\url{#2}}

\bibitem[{\citenamefont{Wegewijs and Nowack}(2005)}]{Wegewijs05}
\bibinfo{author}{\bibfnamefont{M.~R.} \bibnamefont{Wegewijs}} \bibnamefont{and}
  \bibinfo{author}{\bibfnamefont{K.~C.} \bibnamefont{Nowack}},
  \bibinfo{journal}{New J. of Phys.} \textbf{\bibinfo{volume}{7}},
  \bibinfo{pages}{239} (\bibinfo{year}{2005}).

\bibitem[{\citenamefont{Kaat and Flensberg}(2005)}]{Kaat05}
\bibinfo{author}{\bibfnamefont{G.~A.} \bibnamefont{Kaat}} \bibnamefont{and}
  \bibinfo{author}{\bibfnamefont{K.}~\bibnamefont{Flensberg}},
  \bibinfo{journal}{Phys.\ Rev.\ B} \textbf{\bibinfo{volume}{71}},
  \bibinfo{pages}{155408} (\bibinfo{year}{2005}).

\bibitem[{\citenamefont{Koch and von Oppen}(2005{\natexlab{a}})}]{Koch05b}
\bibinfo{author}{\bibfnamefont{J.}~\bibnamefont{Koch}} \bibnamefont{and}
  \bibinfo{author}{\bibfnamefont{F.}~\bibnamefont{von Oppen}},
  \bibinfo{journal}{Phys.\ Rev.\ B} \textbf{\bibinfo{volume}{72}},
  \bibinfo{pages}{113308} (\bibinfo{year}{2005}{\natexlab{a}}).

\bibitem[{\citenamefont{Donarini et~al.}(2006)\citenamefont{Donarini, Grifoni,
  and Richter}}]{Donarini06}
\bibinfo{author}{\bibfnamefont{A.}~\bibnamefont{Donarini}},
  \bibinfo{author}{\bibfnamefont{M.}~\bibnamefont{Grifoni}}, \bibnamefont{and}
  \bibinfo{author}{\bibfnamefont{K.}~\bibnamefont{Richter}},
  \bibinfo{journal}{Phys.\ Rev.\ Lett.} \textbf{\bibinfo{volume}{97}},
  \bibinfo{pages}{166801} (\bibinfo{year}{2006}).

\bibitem[{\citenamefont{Schultz et~al.}(2008)\citenamefont{Schultz, Nunner, and
  von Oppen}}]{Schultz07a}
\bibinfo{author}{\bibfnamefont{M.~G.} \bibnamefont{Schultz}},
  \bibinfo{author}{\bibfnamefont{T.~S.} \bibnamefont{Nunner}},
  \bibnamefont{and} \bibinfo{author}{\bibfnamefont{F.}~\bibnamefont{von
  Oppen}}, \bibinfo{journal}{Phys.\ Rev.\ B} \textbf{\bibinfo{volume}{77}},
  \bibinfo{pages}{075323} (\bibinfo{year}{2008}).

\bibitem[{\citenamefont{Reckermann et~al.}(2008)\citenamefont{Reckermann,
  Leijnse, Wegewijs, and Schoeller}}]{Reckermann08a}
\bibinfo{author}{\bibfnamefont{F.}~\bibnamefont{Reckermann}},
  \bibinfo{author}{\bibfnamefont{M.}~\bibnamefont{Leijnse}},
  \bibinfo{author}{\bibfnamefont{M.~R.} \bibnamefont{Wegewijs}},
  \bibnamefont{and}
  \bibinfo{author}{\bibfnamefont{H.}~\bibnamefont{Schoeller}},
  \bibinfo{journal}{Eur.\ Phys.\ Lett.} \textbf{\bibinfo{volume}{83}},
  \bibinfo{pages}{58001} (\bibinfo{year}{2008}).

\bibitem[{\citenamefont{Koch and von Oppen}(2005{\natexlab{b}})}]{Koch04b}
\bibinfo{author}{\bibfnamefont{J.}~\bibnamefont{Koch}} \bibnamefont{and}
  \bibinfo{author}{\bibfnamefont{F.}~\bibnamefont{von Oppen}},
  \bibinfo{journal}{Phys.\ Rev.\ Lett.} \textbf{\bibinfo{volume}{94}},
  \bibinfo{pages}{206804} (\bibinfo{year}{2005}{\natexlab{b}}).

\bibitem[{\citenamefont{Koch et~al.}(2006{\natexlab{a}})\citenamefont{Koch, von
  Oppen, and Andreev}}]{Koch06}
\bibinfo{author}{\bibfnamefont{J.}~\bibnamefont{Koch}},
  \bibinfo{author}{\bibfnamefont{F.}~\bibnamefont{von Oppen}},
  \bibnamefont{and} \bibinfo{author}{\bibfnamefont{A.~V.}
  \bibnamefont{Andreev}}, \bibinfo{journal}{Phys.\ Rev.\ B}
  \textbf{\bibinfo{volume}{74}}, \bibinfo{pages}{205438}
  (\bibinfo{year}{2006}{\natexlab{a}}).

\bibitem[{\citenamefont{Cornaglia and Grempel}(2005)}]{Cornaglia05b}
\bibinfo{author}{\bibfnamefont{P.~S.} \bibnamefont{Cornaglia}}
  \bibnamefont{and} \bibinfo{author}{\bibfnamefont{D.~R.}
  \bibnamefont{Grempel}}, \bibinfo{journal}{Phys.\ Rev.\ B}
  \textbf{\bibinfo{volume}{71}}, \bibinfo{pages}{245326}
  (\bibinfo{year}{2005}).

\bibitem[{\citenamefont{Koch et~al.}(2006{\natexlab{b}})\citenamefont{Koch,
  Raikh, and von Oppen}}]{Koch05c}
\bibinfo{author}{\bibfnamefont{J.}~\bibnamefont{Koch}},
  \bibinfo{author}{\bibfnamefont{M.~E.} \bibnamefont{Raikh}}, \bibnamefont{and}
  \bibinfo{author}{\bibfnamefont{F.}~\bibnamefont{von Oppen}},
  \bibinfo{journal}{PRL} \textbf{\bibinfo{volume}{96}}, \bibinfo{pages}{056803}
  (\bibinfo{year}{2006}{\natexlab{b}}).

\bibitem[{\citenamefont{Flensberg}(2003)}]{Flensberg03}
\bibinfo{author}{\bibfnamefont{K.}~\bibnamefont{Flensberg}},
  \bibinfo{journal}{Phys.\ Rev.\ B} \textbf{\bibinfo{volume}{68}},
  \bibinfo{pages}{205323} (\bibinfo{year}{2003}).

\bibitem[{\citenamefont{Reckermann et~al.}(2009)\citenamefont{Reckermann,
  Leijnse, and Wegewijs}}]{Reckermann08b}
\bibinfo{author}{\bibfnamefont{F.}~\bibnamefont{Reckermann}},
  \bibinfo{author}{\bibfnamefont{M.}~\bibnamefont{Leijnse}}, \bibnamefont{and}
  \bibinfo{author}{\bibfnamefont{M.~R.} \bibnamefont{Wegewijs}},
  \bibinfo{journal}{Phys. Rev. B} \textbf{\bibinfo{volume}{79}},
  \bibinfo{pages}{075313} (\bibinfo{year}{2009}).

\bibitem[{\citenamefont{Yu et~al.}(2004)\citenamefont{Yu, Keane, Ciszek, Cheng,
  Stewart, Tour, and Natelson}}]{Yu04inel}
\bibinfo{author}{\bibfnamefont{L.~H.} \bibnamefont{Yu}},
  \bibinfo{author}{\bibfnamefont{Z.~K.} \bibnamefont{Keane}},
  \bibinfo{author}{\bibfnamefont{J.~W.} \bibnamefont{Ciszek}},
  \bibinfo{author}{\bibfnamefont{L.}~\bibnamefont{Cheng}},
  \bibinfo{author}{\bibfnamefont{M.~P.} \bibnamefont{Stewart}},
  \bibinfo{author}{\bibfnamefont{J.~M.} \bibnamefont{Tour}}, \bibnamefont{and}
  \bibinfo{author}{\bibfnamefont{D.}~\bibnamefont{Natelson}},
  \bibinfo{journal}{Phys.\ Rev.\ Lett.} \textbf{\bibinfo{volume}{93}},
  \bibinfo{pages}{266802} (\bibinfo{year}{2004}).

\bibitem[{\citenamefont{Natelson}(2006)}]{Natelson06}
\bibinfo{author}{\bibfnamefont{D.}~\bibnamefont{Natelson}},
  \emph{\bibinfo{title}{Handbook of Organic Electronics and Photonics}}
  (\bibinfo{publisher}{American Scientific Publishers},
  \bibinfo{address}{Valencia}, \bibinfo{year}{2006}).

\bibitem[{\citenamefont{Osorio et~al.}(2007{\natexlab{a}})\citenamefont{Osorio,
  O'Neill, Wegewijs, Stuhr-Hansen, Paaske, Bj{\o}rnholm, and van~der
  Zant}}]{Osorio07b}
\bibinfo{author}{\bibfnamefont{E.~A.} \bibnamefont{Osorio}},
  \bibinfo{author}{\bibfnamefont{K.}~\bibnamefont{O'Neill}},
  \bibinfo{author}{\bibfnamefont{M.~R.} \bibnamefont{Wegewijs}},
  \bibinfo{author}{\bibfnamefont{N.}~\bibnamefont{Stuhr-Hansen}},
  \bibinfo{author}{\bibfnamefont{J.}~\bibnamefont{Paaske}},
  \bibinfo{author}{\bibfnamefont{T.}~\bibnamefont{Bj{\o}rnholm}},
  \bibnamefont{and} \bibinfo{author}{\bibfnamefont{H.~S.} \bibnamefont{van~der
  Zant}}, \bibinfo{journal}{Nanolett.} \textbf{\bibinfo{volume}{7}},
  \bibinfo{pages}{3336} (\bibinfo{year}{2007}{\natexlab{a}}).

\bibitem[{\citenamefont{Parks et~al.}(2007)\citenamefont{Parks, Champagne,
  Hutchison, Flores-Torres, Abru{\~n}a, and Ralph}}]{Parks07}
\bibinfo{author}{\bibfnamefont{J.~J.} \bibnamefont{Parks}},
  \bibinfo{author}{\bibfnamefont{A.~R.} \bibnamefont{Champagne}},
  \bibinfo{author}{\bibfnamefont{G.~R.} \bibnamefont{Hutchison}},
  \bibinfo{author}{\bibfnamefont{S.}~\bibnamefont{Flores-Torres}},
  \bibinfo{author}{\bibfnamefont{H.~D.} \bibnamefont{Abru{\~n}a}},
  \bibnamefont{and} \bibinfo{author}{\bibfnamefont{D.~C.} \bibnamefont{Ralph}},
  \bibinfo{journal}{Phys.\ Rev.\ Lett.} \textbf{\bibinfo{volume}{99}},
  \bibinfo{pages}{026601} (\bibinfo{year}{2007}).

\bibitem[{\citenamefont{Pasupathy et~al.}(2005)\citenamefont{Pasupathy, Park,
  Chang, Soldatov, Lebedkin, Bialczak, Grose, Donev, Sethna, Ralph
  et~al.}}]{Pasupathy04}
\bibinfo{author}{\bibfnamefont{A.~N.} \bibnamefont{Pasupathy}},
  \bibinfo{author}{\bibfnamefont{J.}~\bibnamefont{Park}},
  \bibinfo{author}{\bibfnamefont{C.}~\bibnamefont{Chang}},
  \bibinfo{author}{\bibfnamefont{A.~V.} \bibnamefont{Soldatov}},
  \bibinfo{author}{\bibfnamefont{S.}~\bibnamefont{Lebedkin}},
  \bibinfo{author}{\bibfnamefont{R.~C.} \bibnamefont{Bialczak}},
  \bibinfo{author}{\bibfnamefont{J.~E.} \bibnamefont{Grose}},
  \bibinfo{author}{\bibfnamefont{L.~A.~K.} \bibnamefont{Donev}},
  \bibinfo{author}{\bibfnamefont{J.~P.} \bibnamefont{Sethna}},
  \bibinfo{author}{\bibfnamefont{D.~C.} \bibnamefont{Ralph}},
  \bibnamefont{et~al.}, \bibinfo{journal}{Nano\ Lett.}
  \textbf{\bibinfo{volume}{5}}, \bibinfo{pages}{203} (\bibinfo{year}{2005}).

\bibitem[{\citenamefont{Park et~al.}(2000)\citenamefont{Park, Park, Lim,
  Anderson, Alivisatos, and McEuen}}]{Park00}
\bibinfo{author}{\bibfnamefont{H.}~\bibnamefont{Park}},
  \bibinfo{author}{\bibfnamefont{J.}~\bibnamefont{Park}},
  \bibinfo{author}{\bibfnamefont{A.~K.~L.} \bibnamefont{Lim}},
  \bibinfo{author}{\bibfnamefont{E.~H.} \bibnamefont{Anderson}},
  \bibinfo{author}{\bibfnamefont{A.~P.} \bibnamefont{Alivisatos}},
  \bibnamefont{and} \bibinfo{author}{\bibfnamefont{P.~L.}
  \bibnamefont{McEuen}}, \bibinfo{journal}{Nature}
  \textbf{\bibinfo{volume}{407}}, \bibinfo{pages}{57} (\bibinfo{year}{2000}).

\bibitem[{\citenamefont{Osorio et~al.}(2007{\natexlab{b}})\citenamefont{Osorio,
  O'Neill, Stuhr-Hansen, Nielsen, Bj\"ornholm, and van~der Zant}}]{Osorio07a}
\bibinfo{author}{\bibfnamefont{E.~A.} \bibnamefont{Osorio}},
  \bibinfo{author}{\bibfnamefont{K.}~\bibnamefont{O'Neill}},
  \bibinfo{author}{\bibfnamefont{N.}~\bibnamefont{Stuhr-Hansen}},
  \bibinfo{author}{\bibfnamefont{O.~F.} \bibnamefont{Nielsen}},
  \bibinfo{author}{\bibfnamefont{T.}~\bibnamefont{Bj\"ornholm}},
  \bibnamefont{and} \bibinfo{author}{\bibfnamefont{H.~S.} \bibnamefont{van~der
  Zant}}, \bibinfo{journal}{Adv. Mater.} \textbf{\bibinfo{volume}{19}},
  \bibinfo{pages}{281} (\bibinfo{year}{2007}{\natexlab{b}}).

\bibitem[{\citenamefont{Houck et~al.}(2005)\citenamefont{Houck, Labaziewicz,
  Chan, Folk, and Chuang}}]{Houck05}
\bibinfo{author}{\bibfnamefont{A.~A.} \bibnamefont{Houck}},
  \bibinfo{author}{\bibfnamefont{J.}~\bibnamefont{Labaziewicz}},
  \bibinfo{author}{\bibfnamefont{E.~K.} \bibnamefont{Chan}},
  \bibinfo{author}{\bibfnamefont{J.~A.} \bibnamefont{Folk}}, \bibnamefont{and}
  \bibinfo{author}{\bibfnamefont{I.~L.} \bibnamefont{Chuang}},
  \bibinfo{journal}{Nano Lett.} \textbf{\bibinfo{volume}{5}},
  \bibinfo{pages}{1685} (\bibinfo{year}{2005}).

\bibitem[{\citenamefont{Sapmaz et~al.}(2006)\citenamefont{Sapmaz,
  Jarillo-Herrero, Blanter, Dekker, and van~der Zant}}]{Sapmaz05}
\bibinfo{author}{\bibfnamefont{S.}~\bibnamefont{Sapmaz}},
  \bibinfo{author}{\bibfnamefont{P.}~\bibnamefont{Jarillo-Herrero}},
  \bibinfo{author}{\bibfnamefont{Y.~M.} \bibnamefont{Blanter}},
  \bibinfo{author}{\bibfnamefont{C.}~\bibnamefont{Dekker}}, \bibnamefont{and}
  \bibinfo{author}{\bibfnamefont{H.~S.~J.} \bibnamefont{van~der Zant}},
  \bibinfo{journal}{Phys.\ Rev.\ Lett.} \textbf{\bibinfo{volume}{96}},
  \bibinfo{pages}{026801} (\bibinfo{year}{2006}).

\bibitem[{\citenamefont{Sapmaz et~al.}(2003)\citenamefont{Sapmaz, Blanter,
  Gurevich, and van~der Zant}}]{Sapmaz03}
\bibinfo{author}{\bibfnamefont{S.}~\bibnamefont{Sapmaz}},
  \bibinfo{author}{\bibfnamefont{Y.~M.} \bibnamefont{Blanter}},
  \bibinfo{author}{\bibfnamefont{L.}~\bibnamefont{Gurevich}}, \bibnamefont{and}
  \bibinfo{author}{\bibfnamefont{H.~S.~J.} \bibnamefont{van~der Zant}},
  \bibinfo{journal}{Phys.\ Rev.\ B} \textbf{\bibinfo{volume}{67}},
  \bibinfo{pages}{235414} (\bibinfo{year}{2003}).

\bibitem[{\citenamefont{Sazonova et~al.}(2004)\citenamefont{Sazonova, Yaish,
  Ustunel, Roundy, Arias, and McEuen}}]{Sazonova04}
\bibinfo{author}{\bibfnamefont{V.}~\bibnamefont{Sazonova}},
  \bibinfo{author}{\bibfnamefont{Y.}~\bibnamefont{Yaish}},
  \bibinfo{author}{\bibfnamefont{H.}~\bibnamefont{Ustunel}},
  \bibinfo{author}{\bibfnamefont{D.}~\bibnamefont{Roundy}},
  \bibinfo{author}{\bibfnamefont{T.~A.} \bibnamefont{Arias}}, \bibnamefont{and}
  \bibinfo{author}{\bibfnamefont{P.~L.} \bibnamefont{McEuen}},
  \bibinfo{journal}{Nature} \textbf{\bibinfo{volume}{431}},
  \bibinfo{pages}{284} (\bibinfo{year}{2004}).

\bibitem[{\citenamefont{Mitra et~al.}(2004)\citenamefont{Mitra, Aleiner, and
  Millis}}]{Mitra04b}
\bibinfo{author}{\bibfnamefont{A.}~\bibnamefont{Mitra}},
  \bibinfo{author}{\bibfnamefont{I.}~\bibnamefont{Aleiner}}, \bibnamefont{and}
  \bibinfo{author}{\bibfnamefont{A.~J.} \bibnamefont{Millis}},
  \bibinfo{journal}{Phys.\ Rev.\ B} \textbf{\bibinfo{volume}{69}},
  \bibinfo{pages}{245302} (\bibinfo{year}{2004}).

\bibitem[{\citenamefont{Schleser et~al.}(2005)\citenamefont{Schleser, Ihn, Ruh,
  Ensslin, Tews, Pfannkuche, Driscoll, and Gossard}}]{Schleser05}
\bibinfo{author}{\bibfnamefont{R.}~\bibnamefont{Schleser}},
  \bibinfo{author}{\bibfnamefont{T.}~\bibnamefont{Ihn}},
  \bibinfo{author}{\bibfnamefont{E.}~\bibnamefont{Ruh}},
  \bibinfo{author}{\bibfnamefont{K.}~\bibnamefont{Ensslin}},
  \bibinfo{author}{\bibfnamefont{M.}~\bibnamefont{Tews}},
  \bibinfo{author}{\bibfnamefont{D.}~\bibnamefont{Pfannkuche}},
  \bibinfo{author}{\bibfnamefont{D.~C.} \bibnamefont{Driscoll}},
  \bibnamefont{and} \bibinfo{author}{\bibfnamefont{A.~C.}
  \bibnamefont{Gossard}}, \bibinfo{journal}{Phys.\ Rev.\ Lett.}
  \textbf{\bibinfo{volume}{94}}, \bibinfo{pages}{206805}
  (\bibinfo{year}{2005}).

\bibitem[{\citenamefont{Golovach and Loss}(2004)}]{Golovach04}
\bibinfo{author}{\bibfnamefont{V.~N.} \bibnamefont{Golovach}} \bibnamefont{and}
  \bibinfo{author}{\bibfnamefont{D.}~\bibnamefont{Loss}},
  \bibinfo{journal}{Phys.\ Rev.\ B} \textbf{\bibinfo{volume}{69}},
  \bibinfo{pages}{245327} (\bibinfo{year}{2004}).

\bibitem[{\citenamefont{Elste and Timm}(2007)}]{Elste07}
\bibinfo{author}{\bibfnamefont{F.}~\bibnamefont{Elste}} \bibnamefont{and}
  \bibinfo{author}{\bibfnamefont{C.}~\bibnamefont{Timm}},
  \bibinfo{journal}{Phys.\ Rev.\ B} \textbf{\bibinfo{volume}{75}},
  \bibinfo{pages}{195341} (\bibinfo{year}{2007}).

\bibitem[{\citenamefont{L\"uffe et~al.}(2008)\citenamefont{L\"uffe, Koch, and
  von Oppen}}]{Luffe07}
\bibinfo{author}{\bibfnamefont{M.~C.} \bibnamefont{L\"uffe}},
  \bibinfo{author}{\bibfnamefont{J.}~\bibnamefont{Koch}}, \bibnamefont{and}
  \bibinfo{author}{\bibfnamefont{F.}~\bibnamefont{von Oppen}},
  \bibinfo{journal}{Phys.\ Rev.\ B} \textbf{\bibinfo{volume}{77}},
  \bibinfo{pages}{125306} (\bibinfo{year}{2008}).

\bibitem[{\citenamefont{K\"onig et~al.}(1998)\citenamefont{K\"onig, Schoeller,
  and Sch\"on}}]{Koenig98}
\bibinfo{author}{\bibfnamefont{J.}~\bibnamefont{K\"onig}},
  \bibinfo{author}{\bibfnamefont{H.}~\bibnamefont{Schoeller}},
  \bibnamefont{and} \bibinfo{author}{\bibfnamefont{G.}~\bibnamefont{Sch\"on}},
  \bibinfo{journal}{Phys.\ Rev.\ B} \textbf{\bibinfo{volume}{58}},
  \bibinfo{pages}{7882} (\bibinfo{year}{1998}).

\bibitem[{\citenamefont{K\"onig}(1999)}]{Koenig99}
\bibinfo{author}{\bibfnamefont{J.}~\bibnamefont{K\"onig}}, Ph.D. thesis
  (\bibinfo{year}{1999}).

\bibitem[{\citenamefont{Kubala and K\"{o}nig}(2006)}]{Kubala06}
\bibinfo{author}{\bibfnamefont{B.}~\bibnamefont{Kubala}} \bibnamefont{and}
  \bibinfo{author}{\bibfnamefont{J.}~\bibnamefont{K\"{o}nig}},
  \bibinfo{journal}{PRB} \textbf{\bibinfo{volume}{73}}, \bibinfo{pages}{195316}
  (\bibinfo{year}{2006}).

\bibitem[{\citenamefont{Timm}(2008)}]{Timm08}
\bibinfo{author}{\bibfnamefont{C.}~\bibnamefont{Timm}},
  \bibinfo{journal}{Phys.\ Rev.\ B} \textbf{\bibinfo{volume}{77}},
  \bibinfo{pages}{195416} (\bibinfo{year}{2008}).

\bibitem[{\citenamefont{Schoeller and Sch\"on}(1994)}]{Schoeller94}
\bibinfo{author}{\bibfnamefont{H.}~\bibnamefont{Schoeller}} \bibnamefont{and}
  \bibinfo{author}{\bibfnamefont{G.}~\bibnamefont{Sch\"on}},
  \bibinfo{journal}{Phys.\ Rev.\ B} \textbf{\bibinfo{volume}{50}},
  \bibinfo{pages}{18436} (\bibinfo{year}{1994}).

\bibitem[{\citenamefont{K\"onig et~al.}(1997)\citenamefont{K\"onig, Schoeller,
  and Sch\"on}}]{Koenig97}
\bibinfo{author}{\bibfnamefont{J.}~\bibnamefont{K\"onig}},
  \bibinfo{author}{\bibfnamefont{H.}~\bibnamefont{Schoeller}},
  \bibnamefont{and} \bibinfo{author}{\bibfnamefont{G.}~\bibnamefont{Sch\"on}},
  \bibinfo{journal}{Phys.\ Rev.\ Lett.} \textbf{\bibinfo{volume}{78}},
  \bibinfo{pages}{4482} (\bibinfo{year}{1997}).

\bibitem[{\citenamefont{Schoeller}(2009)}]{Schoeller08-rtrg}
\bibinfo{author}{\bibfnamefont{H.}~\bibnamefont{Schoeller}},
  \bibinfo{journal}{Eur. Phys. Journ. Special Topics}
  \textbf{\bibinfo{volume}{168}}, \bibinfo{pages}{179} (\bibinfo{year}{2009}).

\bibitem[{\citenamefont{Thielmann et~al.}(2005)\citenamefont{Thielmann,
  Hettler, K\"onig, and Sch\"on}}]{Thielmann05}
\bibinfo{author}{\bibfnamefont{A.}~\bibnamefont{Thielmann}},
  \bibinfo{author}{\bibfnamefont{M.~H.} \bibnamefont{Hettler}},
  \bibinfo{author}{\bibfnamefont{J.}~\bibnamefont{K\"onig}}, \bibnamefont{and}
  \bibinfo{author}{\bibfnamefont{G.}~\bibnamefont{Sch\"on}},
  \bibinfo{journal}{Phys.\ Rev.\ Lett.} \textbf{\bibinfo{volume}{95}},
  \bibinfo{pages}{146806} (\bibinfo{year}{2005}).

\bibitem[{\citenamefont{Weymann et~al.}(2005)\citenamefont{Weymann, K{\"o}nig,
  Martinek, Barnas, and Sch{\"o}n}}]{Weymann05}
\bibinfo{author}{\bibfnamefont{I.}~\bibnamefont{Weymann}},
  \bibinfo{author}{\bibfnamefont{J.}~\bibnamefont{K{\"o}nig}},
  \bibinfo{author}{\bibfnamefont{J.}~\bibnamefont{Martinek}},
  \bibinfo{author}{\bibfnamefont{J.}~\bibnamefont{Barnas}}, \bibnamefont{and}
  \bibinfo{author}{\bibfnamefont{G.}~\bibnamefont{Sch{\"o}n}},
  \bibinfo{journal}{Phys.\ Rev.\ B} \textbf{\bibinfo{volume}{72}},
  \bibinfo{pages}{115334} (\bibinfo{year}{2005}).

\bibitem[{\citenamefont{Korb et~al.}(2007)\citenamefont{Korb, Reininghaus,
  Schoeller, and K\"onig}}]{Korb07}
\bibinfo{author}{\bibfnamefont{T.}~\bibnamefont{Korb}},
  \bibinfo{author}{\bibfnamefont{F.}~\bibnamefont{Reininghaus}},
  \bibinfo{author}{\bibfnamefont{H.}~\bibnamefont{Schoeller}},
  \bibnamefont{and} \bibinfo{author}{\bibfnamefont{J.}~\bibnamefont{K\"onig}},
  \bibinfo{journal}{Phys.\ Rev.\ B} \textbf{\bibinfo{volume}{76}},
  \bibinfo{pages}{165316} (\bibinfo{year}{2007}).

\bibitem[{\citenamefont{Blum}(1996)}]{Blum_book}
\bibinfo{author}{\bibfnamefont{K.}~\bibnamefont{Blum}},
  \emph{\bibinfo{title}{Density Matrix Theory and Applications}}
  (\bibinfo{publisher}{Plenum Press}, \bibinfo{address}{New York},
  \bibinfo{year}{1996}).

\bibitem[{\citenamefont{Becker and Pfannkuche}(2008)}]{Becker07}
\bibinfo{author}{\bibfnamefont{D.}~\bibnamefont{Becker}} \bibnamefont{and}
  \bibinfo{author}{\bibfnamefont{D.}~\bibnamefont{Pfannkuche}},
  \bibinfo{journal}{Phys.\ Rev.\ B} \textbf{\bibinfo{volume}{77}},
  \bibinfo{pages}{205307} (\bibinfo{year}{2008}).

\bibitem[{\citenamefont{Nowack and Wegewijs}()}]{Nowack05}
\bibinfo{author}{\bibfnamefont{K.~C.} \bibnamefont{Nowack}} \bibnamefont{and}
  \bibinfo{author}{\bibfnamefont{M.~R.} \bibnamefont{Wegewijs}},
  \bibinfo{note}{cond-mat/0506552}.

\bibitem[{\citenamefont{Aghassi et~al.}(2008)\citenamefont{Aghassi, Hettler,
  and Sch\"on}}]{Aghassi08}
\bibinfo{author}{\bibfnamefont{J.}~\bibnamefont{Aghassi}},
  \bibinfo{author}{\bibfnamefont{M.}~\bibnamefont{Hettler}}, \bibnamefont{and}
  \bibinfo{author}{\bibfnamefont{G.}~\bibnamefont{Sch\"on}},
  \bibinfo{journal}{APL} \textbf{\bibinfo{volume}{92}}, \bibinfo{pages}{202101}
  (\bibinfo{year}{2008}).

\bibitem[{\citenamefont{Galperin et~al.}(2007)\citenamefont{Galperin, Ratner,
  and Nitzan}}]{Galperin07}
\bibinfo{author}{\bibfnamefont{M.}~\bibnamefont{Galperin}},
  \bibinfo{author}{\bibfnamefont{M.~A.} \bibnamefont{Ratner}},
  \bibnamefont{and} \bibinfo{author}{\bibfnamefont{A.}~\bibnamefont{Nitzan}},
  \bibinfo{journal}{J. Phys.: Condens. Matter} \textbf{\bibinfo{volume}{19}},
  \bibinfo{pages}{103201} (\bibinfo{year}{2007}).

\bibitem[{\citenamefont{Braig and Flensberg}(2003)}]{Braig03a}
\bibinfo{author}{\bibfnamefont{S.}~\bibnamefont{Braig}} \bibnamefont{and}
  \bibinfo{author}{\bibfnamefont{K.}~\bibnamefont{Flensberg}},
  \bibinfo{journal}{Phys.\ Rev.\ B} \textbf{\bibinfo{volume}{68}},
  \bibinfo{pages}{205324} (\bibinfo{year}{2003}).

\bibitem[{\citenamefont{Pistolesi and Labarthe}(2007)}]{Pistolesi07}
\bibinfo{author}{\bibfnamefont{F.}~\bibnamefont{Pistolesi}} \bibnamefont{and}
  \bibinfo{author}{\bibfnamefont{S.}~\bibnamefont{Labarthe}},
  \bibinfo{journal}{Phys.\ Rev.\ B} \textbf{\bibinfo{volume}{76}},
  \bibinfo{pages}{165317} (\bibinfo{year}{2007}).

\bibitem[{\citenamefont{Gardiner}(1991)}]{Gardiner_noisebook}
\bibinfo{author}{\bibfnamefont{C.~W.} \bibnamefont{Gardiner}},
  \emph{\bibinfo{title}{Quantum noise}} (\bibinfo{publisher}{Springer-Verlag},
  \bibinfo{address}{Berlin}, \bibinfo{year}{1991}).

\bibitem[{\citenamefont{Leggett et~al.}(1987)\citenamefont{Leggett,
  Chakravarty, Dorsey, Fisher, Garg, and Zwerger}}]{Legget87}
\bibinfo{author}{\bibfnamefont{A.~J.} \bibnamefont{Leggett}},
  \bibinfo{author}{\bibfnamefont{S.}~\bibnamefont{Chakravarty}},
  \bibinfo{author}{\bibfnamefont{A.~T.} \bibnamefont{Dorsey}},
  \bibinfo{author}{\bibfnamefont{M.~P.~A.} \bibnamefont{Fisher}},
  \bibinfo{author}{\bibfnamefont{A.}~\bibnamefont{Garg}}, \bibnamefont{and}
  \bibinfo{author}{\bibfnamefont{W.}~\bibnamefont{Zwerger}},
  \bibinfo{journal}{Rev.\ Mod.\ Phys.} \textbf{\bibinfo{volume}{59}},
  \bibinfo{pages}{1} (\bibinfo{year}{1987}).

\bibitem[{\citenamefont{K\"onig}(1995)}]{Koenig_master}
\bibinfo{author}{\bibfnamefont{J.}~\bibnamefont{K\"onig}}, Master's thesis,
  \bibinfo{school}{Karlsruhe} (\bibinfo{year}{1995}).

\end{thebibliography}

\end{document}